\documentclass[11pt]{article}
\usepackage[preprint]{acl}
\usepackage{times}
\usepackage{latexsym}
\usepackage[T1]{fontenc}
\usepackage[utf8]{inputenc}
\usepackage{microtype}
\usepackage{inconsolata}
\usepackage{graphicx}
\usepackage{xcolor}
\usepackage{amsmath,amssymb,bbm}
\usepackage{xspace}
\usepackage{pifont}
\usepackage{colortbl}

\definecolor{navy}{rgb}{0.0, 0.0, 0.5}
\definecolor{pinklight}{RGB}{240, 181, 211}
\definecolor{pinkdark}{RGB}{168, 75, 122}
\definecolor{checkgreen}{RGB}{34, 139, 34}
\definecolor{crossred}{RGB}{178, 34, 34}
\definecolor{ourscolor}{RGB}{230, 230, 240}

\usepackage{tcolorbox}
\tcbuselibrary{theorems,breakable}
\usepackage{hyperref}
\usepackage{cleveref}
\usepackage{booktabs}
\usepackage{adjustbox}
\usepackage{multirow}
\usepackage{algorithm}
\usepackage{algorithmic}

\definecolor{teallight}{RGB}{224, 247, 250}
\definecolor{tealdark}{RGB}{0, 128, 128}

\newtcbtheorem[auto counter,
crefname={example}{Example},
Crefname={Example}{Example}]
{prompt}{Example}
{colback=pinklight!20, colframe=pinkdark, left=.02in, right=.02in,bottom=.02in, top=.02in}{prompt}

\newcommand{\our}{\textsc{SoundBreak}\xspace}
\newcommand{\ourbold}{\textbf{\textsc{SoundBreak}}\xspace}
\newcommand{\cmark}{{\color{checkgreen}\ding{51}}}
\newcommand{\xmark}{{\color{crossred}\ding{55}}}

\title{\ourbold: A Systematic Study of Audio-Only Adversarial Attacks on Trimodal Models}

\author{
 \textbf{Aafiya Hussain\textsuperscript{$\heartsuit$}},
 \textbf{Gaurav Srivastava\textsuperscript{$\heartsuit$}},
  \textbf{Alvi Ishmam\textsuperscript{$\heartsuit$}},
  \textbf{Zaber Hakim\textsuperscript{$\heartsuit$}},
 \textbf{Chris Thomas\textsuperscript{$\heartsuit$}} 
\\
 \textsuperscript{$\heartsuit$}Department of Computer Science, Virginia Tech, USA,
\\
  \textsuperscript{$\heartsuit$}\texttt{(\href{aafiyahussain@vt.edu}{aafiyahussain}, \href{gks@vt.edu}{gks}, \href{alvi@vt.edu}{alvi}, \href{zaberhakim666@vt.edu}{zaberhakim666}, \href{christhomas@vt.edu}{christhomas})@vt.edu}
  \\
}

\begin{document}

\maketitle
\setcounter{tocdepth}{-1}

\begin{abstract}
Multimodal foundation models that integrate audio, vision, and language achieve strong performance on reasoning and generation tasks, yet their robustness to adversarial manipulation remains poorly understood. We study a realistic and underexplored threat model: \textbf{untargeted, audio-only adversarial attacks} on trimodal audio–video–language models. We analyze six complementary attack objectives that target different stages of multimodal processing, including audio encoder representations, cross-modal attention, hidden states, and output likelihoods. Across three state-of-the-art models and multiple benchmarks, we show that audio-only perturbations can induce severe multimodal failures, achieving up to \textbf{96\% attack success rate.} We further show that attacks can be successful at low perceptual distortions (LPIPS $\leq 0.08$, SI-SNR $\geq 0$) and benefit more from extended optimization than increased data scale. Transferability across models and encoders remains limited, while speech recognition systems such as Whisper primarily respond to perturbation magnitude, achieving \textbf{$>$97\% attack success} under severe distortion. These results expose a previously overlooked single-modality attack surface in multimodal systems and motivate defenses that enforce cross-modal consistency.
\end{abstract}

\section{Introduction}
\label{sec:intro}
\begin{figure}[ht]
    \centering
    \includegraphics[width=\linewidth]{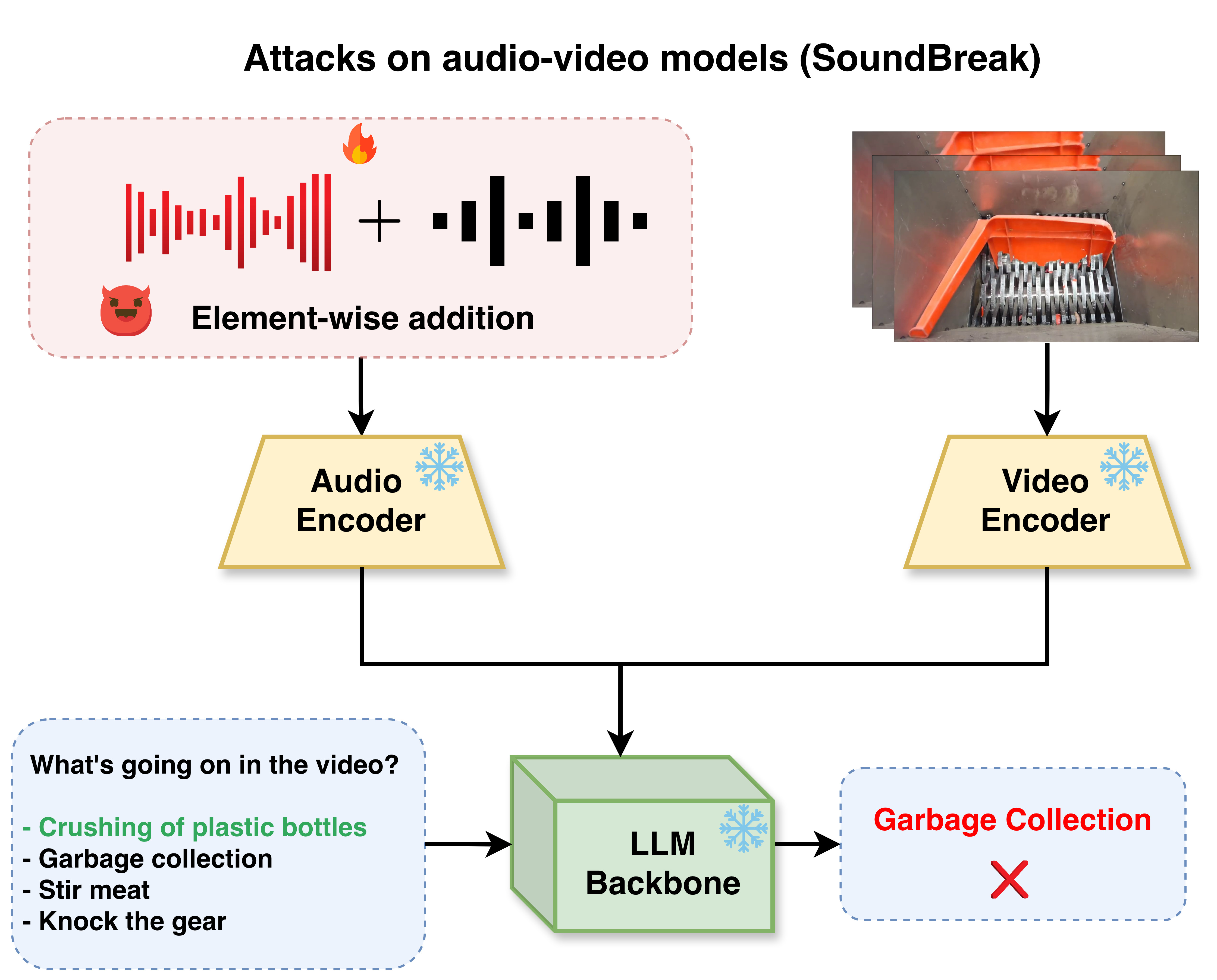}
    \caption{Audio-only adversarial attacks on audio–video–language models. An additive perturbation applied solely to the audio stream propagates through the model, resulting in incorrect outputs.}
    \label{fig:concept}
\end{figure}
Recent progress in multimodal large language models has intensified research on their susceptibility to adversarial attacks across multiple modalities~\cite{liu2025survey}. While these models demonstrate strong capabilities in integrating audio, vision, and language, prior work has revealed critical vulnerabilities in CLIP-based and vision-language systems~\cite{zhang2025qava, mei2025veattack, xu2024shadowcast}, with growing interest in audio-domain attacks~\cite{raina2024muting}. Despite these advances, robustness against adversarial manipulation remains largely unexplored in settings where the attacker controls only a single modality within a multimodal input pipeline.

Adversarial attacks on multimodal models broadly fall into four categories: \textit{prompt injection}, \textit{jailbreaks}, \textit{data poisoning}, and \textit{adversarial perturbations}~\cite{liu2025survey, jiang2025survey, shayegani2023jailbreak, lu2025adversarial}. Jailbreak attacks bypass safety mechanisms through prompt engineering~\cite{yi2024jailbreakattacksdefenseslarge, liu2024jailbreak} or gradient-based optimization~\cite{wang2024whiteboxmultimodaljailbreakslarge}, while prompt injection embeds malicious instructions directly into inputs such as images or audio~\cite{bagdasaryan2023abusingimagessoundsindirect}. Data poisoning and backdoor attacks compromise training pipelines~\cite{walmer2022dual}. Among these, adversarial perturbations pose a particularly unique cross-modal threat: by modifying a single input modality, they can indirectly influence model reasoning across all modalities without explicitly altering them~\cite{kang2025traptargetedredirectingagentic}. This attack surface remains insufficiently understood.

Existing work has predominantly focused on unimodal or bimodal settings. Vision-language attacks demonstrate that carefully crafted image perturbations can manipulate model outputs across downstream tasks~\cite{zhang2025qava, mei2025veattack, hao2025exploring, zhang2025modality, zhang2022towards}. Separately, audio-domain attacks show that speech recognition and translation systems can be silenced, redirected, or manipulated through acoustic adversarial examples~\cite{raina2024muting, ma2025universal, sadasivan2025attacker, liu2023privacy}. However, these approaches treat modalities largely in isolation, failing to capture the complex cross-modal dependencies inherent to modern multimodal foundation models.

A small number of multimodal attack studies rely on simultaneous manipulation of multiple input channels~\cite{mustakim2025watchlistenunderstandmislead, wang2025manipulating, tian2021can} or require access to training data for backdoor insertion~\cite{walmer2022dual, han2024backdooring, liang2024badclip, yu2025backdoormbti, liang2025vl}. These assumptions limit practical relevance: coordinating attacks across multiple modalities is technically challenging and easily detectable, while training-time attacks assume access to training data. Moreover, targeted attacks that aim to induce specific malicious outputs often introduce conspicuous behavioral anomalies that can be flagged by monitoring systems.

In this work, we investigate a more realistic and underexplored threat model: \textbf{untargeted, audio-only adversarial attacks on audio-video-language models}. We emphasize that our attacks are not per-sample optimizations, but shared perturbations learned through optimization over training data. We demonstrate that an attacker who controls only the audio channel can systematically degrade multimodal reasoning through gradient-based optimization, without modifying visual or textual inputs. This threat model is particularly concerning because audio manipulation is easier to deploy in real-world settings via compromised microphones, environmental speakers, or transmission channels, while being significantly harder to detect than visual perturbations~\cite{choi2024ghost}.

Our contributions are threefold. \textbf{\textit{First,}} we propose six complementary audio-based adversarial objectives that target different stages of multimodal processing, including encoder representations, attention allocation, hidden states, and output likelihoods. \textbf{\textit{Second,}} we conduct extensive evaluations across three state-of-the-art audio-visual models on standardized multimodal benchmarks, revealing model-specific vulnerabilities and limited cross-model transferability. \textbf{\textit{Third,}} we analyze how attack effectiveness depends on training duration, data efficiency, perturbation magnitude, and perceptual distortion, yielding practical insights into both attack construction and defensive design.

\section{Related Work}
\label{sec:related}

\paragraph{Foundations of Adversarial Attacks.}
The discovery of adversarial examples fundamentally challenged prevailing assumptions about neural network robustness~\cite{szegedy2013intriguing, nguyen2015deep}. Early investigations highlighted the structural fragility of deep models under small, adversarially chosen perturbations~\cite{papernot2016limitations}, while the Fast Gradient Sign Method demonstrated how approximately linear behavior in high-dimensional spaces gives rise to these vulnerabilities~\cite{goodfellow2014explaining, goodfellow2015explainingharnessingadversarialexamples}. Subsequent work introduced increasingly precise optimization-based attacks, ranging from DeepFool's geometric formulation~\cite{moosavi2016deepfool} to the Carlini-Wagner attack~\cite{carlini2017towards}, establishing that adversarial robustness cannot be achieved through superficial defenses alone. More recently, these principles have been successfully extended to large language models via projected gradient descent on relaxed input representations, enabling efficient and scalable attacks in previously intractable settings~\cite{geisler2024attacking}.

\paragraph{Vision-Centric Multimodal Attacks.}
With the rise of vision-language models, adversarial research has increasingly focused on vulnerabilities at the vision-text interface. Query-agnostic attacks demonstrate that a single adversarial image can generalize across multiple downstream questions by disrupting vision-language alignment~\cite{zhang2025qava}. Related work shows that targeting only the vision encoder suffices to degrade performance across diverse tasks, highlighting encoder representations as a critical attack surface~\cite{mei2025veattack}. Scenario-aware and multi-loss attacks further amplify these effects, achieving high jailbreak success rates on commercial systems~\cite{hao2025exploring}. Scalability has been addressed through self-supervised pretraining strategies that enable transferable any-to-any attacks across tasks and prompts~\cite{zhang2025anyattack}. However, these approaches primarily rely on visual perturbations and do not examine how vulnerabilities propagate across more than two modalities.

\paragraph{Audio-Domain Adversarial Techniques.}
The audio modality introduces distinct adversarial challenges due to temporal structure, perceptual constraints, and physical-world variability. Universal acoustic segments can suppress or terminate speech recognition by mimicking control tokens~\cite{raina2024muting}, while targeted perturbations enable conditional manipulation based on speaker identity or content~\cite{ma2025universal}. Over-the-air attacks demonstrate that adversarial audio can remain effective despite environmental noise and reverberation~\cite{sadasivan2025attacker}. Recent studies extend these ideas to speech translation systems, showing that both imperceptible perturbations and adversarial music can induce malicious translations~\cite{liu2025exploiting}. Defense mechanisms such as SPIRIT propose activation patching to mitigate such attacks, but also expose the trade-off between robustness and model utility~\cite{djanibekov2025spirit}. AdvWave further demonstrates that gradient shattering in audio models can be overcome through dual-phase optimization while preserving perceptual naturalness~\cite{kangadvwave}. Notably, these works focus primarily on speech-centric models and do not address multimodal reasoning.

\paragraph{Audio-Only Attacks in Multimodal Systems.}
Table~\ref{tab:related_work_comparison} summarizes how existing adversarial attacks differ in their threat models and assumptions. Prior work largely focuses on vision-language settings, unimodal speech models, or attacks that require coordinated manipulation of multiple modalities. As a result, these approaches either assume access to multiple input channels, target specific outputs, or operate outside truly multimodal reasoning settings. In contrast, \our is the first to study \emph{untargeted, audio-only attacks} in trimodal audio-video-language models under a realistic threat model, while providing systematic analysis of encoder-space, attention-based, and hidden-state level vulnerabilities. This positioning highlights a previously unexplored attack surface in multimodal systems.

\begin{table*}[t]
\centering
\adjustbox{max width=\textwidth}{
\begin{tabular}{lccccccc}
\toprule
 & \textbf{QAVA}
 & \textbf{VEAttack}
 & \textbf{Muting Whisper}
 & \textbf{AdvWave}
 & \textbf{Multimodal Attacks}
 & \cellcolor{ourscolor}\textbf{\our} \\
\midrule
\textbf{Targets Audio Modality}
  & \xmark & \xmark & \cmark & \cmark & \cmark & \cellcolor{ourscolor}\cmark \\

\textbf{Single-Modality Control}
  & \xmark & \cmark & \cmark & \cmark & \xmark & \cellcolor{ourscolor}\cmark \\

\textbf{Trimodal Setting (A+V+L)}
  & \xmark & \xmark & \xmark & \xmark & \cmark & \cellcolor{ourscolor}\cmark \\

\textbf{Untargeted Attack Objective}
  & \cmark & \cmark & \cmark & \xmark & \xmark & \cellcolor{ourscolor}\cmark \\

\textbf{Query-Agnostic / Task-Agnostic}
  & \cmark & \cmark & \cmark & \xmark & \xmark & \cellcolor{ourscolor}\cmark \\

\textbf{Realistic Threat Model}
  & $\triangle$ & $\triangle$ & \cmark & $\triangle$ & \xmark & \cellcolor{ourscolor}\cmark \\

\textbf{Attack Interpretability Analysis}
  & \xmark & \xmark & \xmark & $\triangle$ & \xmark & \cellcolor{ourscolor}\cmark \\

\bottomrule
\end{tabular}
}
\caption{Comparison of prior adversarial attack methods on multimodal models. Existing work largely focuses on vision-language or unimodal settings, targeted objectives, or simultaneous multi-modal manipulation. \our uniquely studies untargeted, audio-only attacks in trimodal models with systematic analysis of encoder-space, attention-based, and hidden-state vulnerabilities, as well as transferability across models and encoders. $\triangle$ denotes partial support. Multimodal Attacks refer to the attack from \cite{mustakim2025watchlistenunderstandmislead}.}
\label{tab:related_work_comparison}
\end{table*}

\section{\our Setup}
\label{sec:method}

\subsection{Problem Formulation}
\label{sec:problem}

Let $M_{\theta}:\mathcal{X}_a\times\mathcal{X}_v\times\mathcal{Q}\to\mathcal{P}(\mathcal{A})$ denote an audio-video model that consumes a natural-language question $q\in\mathcal{Q}$ together with synchronized audio $x_a\in\mathcal{X}_a$ (of length $T$) and video $x_v\in\mathcal{X}_v$, and returns a distribution $p(\cdot\mid x_a,x_v,q;\theta)$ over answers, where $\theta$ represents the model parameters. We study a white-box, audio-only additive adversary that seeks to alter the model's answer in an \emph{untargeted} fashion. The adversary injects a perturbation $\delta\in\mathbb{R}^T$ to produce $\tilde{x}_a=x_a+\delta$ subject to a feasibility set
\begin{equation}
\mathcal{C} = \{\delta : \|\delta\|_\infty \le \varepsilon\},
\end{equation}
which captures the $\ell_\infty$ norm constraint with $\varepsilon$ denoting the attack budget.

Under the white-box assumption, the attacker has full access to $\theta$, model outputs, and internal differentiable representations. The attacker may therefore evaluate gradients $\nabla_{\delta}\mathcal{L}^{(i)}$ for each chosen loss index $i\in\{1,\dots,k\}$. We define the iterative refinement of the additive perturbation by
\begin{equation}
\begin{split}
\delta^{(k+1)} =
\Pi_{\mathcal{C}}\Big(
    \delta^{(k)} - \eta\, \nabla_{\delta}\,
    \mathcal{L}^{(i)}\big(
        p(\cdot \mid x_a + \delta^{(k)}, \\
        x_v, q; \theta),
        p(\cdot \mid x_a, x_v, q; \theta)
    \big)
\Big),
\end{split}
\end{equation}
where $\delta^{(0)} \in \mathcal{C}$, $\Pi_{\mathcal{C}}$ denotes projection onto the feasible set $\mathcal{C}$, and $\eta>0$ is the step size. After $K$ iterations, the attack uses $\tilde{x}_a=x_a+\delta^{(K)}$ as the perturbed audio input. We further evaluate this attack on other models in a black-box setting using $\tilde{x}_a$ as the new audio input.

\subsection{Attack Methodology}
\label{sec:attacks}

We design our attack methodology to systematically stress-test trimodal audio-video-language models under deviations in the audio channel while keeping visual and textual inputs fixed. We employ a diverse set of adversarial losses that probe complementary stages of the audio-to-output pipeline, including representation learning, cross-modal interaction, and internal attention dynamics. This multi-view formulation allows us to disentangle which computational components are most susceptible to audio-only perturbations and how failures propagate across modalities. Lastly, all attacks optimize a shared audio perturbation rather than per-sample adversarial noise, enabling analysis of systematic model vulnerabilities. Detailed mathematical formulations and algorithms are in appendix ~\ref{appendix:loss_formulation}, ~\ref{appendix:algorithm}.

\subsubsection{Negative Language Modeling Loss}
A direct way to induce adversarial failure is to reduce the model's confidence in its original answer under perturbed audio input. We implement this by defining an objective derived from the language modeling loss, which serves as a scalar proxy for answer likelihood in multimodal generation.

Let $a^\star \in \mathcal{A}$ denote the ground-truth answer associated with input $(x_a, x_v, q)$. The standard language modeling loss under perturbed audio $\tilde{x}_a = x_a + \delta$ is given by
\begin{equation}
\begin{split}
\mathcal{L}_{\text{LM}}\big(
p(\cdot \mid \tilde{x}_a, x_v, q; \theta), a^\star
\big)
= \\
- \log p(a^\star \mid \tilde{x}_a, x_v, q; \theta).
\end{split}
\end{equation}
To explicitly suppress the probability assigned to the correct answer, we define the adversarial objective as
\begin{equation}
\mathcal{L}_{\text{negLM}} = -\mathcal{L}_{\text{LM}}.
\end{equation}
This formulation is motivated by recent work showing that directly optimizing inputs against the language modeling objective is an effective mechanism for attacking large language models using projected gradient descent~\cite{geisler2025attackinglargelanguagemodels}.

\subsubsection{Encoder-Based Cosine Similarity Loss}
To directly probe vulnerabilities in audio representation learning, we target the audio encoder rather than downstream language components. Let $f_a : \mathcal{X}_a \rightarrow \mathbb{R}^d$ denote the audio encoder of $M_\theta$, which maps an input waveform to a $d$-dimensional embedding. We define an encoder-space adversarial objective that minimizes the cosine similarity between the embeddings of clean and perturbed audio:
\begin{equation}
\mathcal{L}^{(\cos)}(\tilde{x}_a, x_a)
= \frac{f_a(\tilde{x}_a) \cdot f_a(x_a)}{\lVert f_a(\tilde{x}_a) \rVert_2 \, \lVert f_a(x_a) \rVert_2}.
\end{equation}
This formulation is inspired by encoder-only adversarial attacks in vision-language models, which demonstrate that perturbing modality-specific encoders can induce downstream failure without relying on task supervision~\cite{mei2025veattack}.

\subsubsection{Vision Attention Suppression Loss}
Attention mechanisms play a central role in multimodal models by mediating cross-modal alignment. Prior work has shown that adversarial manipulation of attention patterns can induce failures such as jailbreaking and hallucination~\cite{wang2025attentionvisionlanguagemodel,wang2024attngcgenhancingjailbreakingattacks}. We design an audio-only objective that reduces the model's reliance on visual evidence by suppressing attention allocated to vision tokens.

Let $A_{l,h,t}(x_a + \delta, x_v, q; \theta)$ denote the attention logit at layer $l$, head $h$, and target token position $t$. Let $\mathcal{T}_v$ denote the set of token indices associated with the vision modality. We aggregate the total attention mass assigned to vision tokens as
\begin{equation}
S_v(\delta) = \sum_{l=1}^{L} \sum_{h=1}^{H} \sum_{t \in \mathcal{T}_v}
A_{l,h,t}(x_a + \delta, x_v, q; \theta).
\end{equation}
The adversarial objective is
\begin{equation}
\mathcal{L}^{(\text{visionatt})}(\delta) = S_v(\delta),
\end{equation}
which is minimized during optimization to suppress visual grounding.

\subsubsection{Audio Attention Amplification Loss}
Complementary to suppressing visual attention, we consider an objective that explicitly amplifies the model's reliance on the perturbed audio stream. Let $\mathcal{T}_a$ denote the set of token indices corresponding to the audio modality. We aggregate the total attention mass assigned to audio tokens as
\begin{equation}
S_a(\delta) = \sum_{l=1}^{L} \sum_{h=1}^{H} \sum_{t \in \mathcal{T}_a}
A_{l,h,t}(x_a + \delta, x_v, q; \theta).
\end{equation}
To encourage the model to over-weight the attacked audio channel, we define
\begin{equation}
\mathcal{L}^{(\text{audioatt})}(\delta) = - S_a(\delta),
\end{equation}
which is minimized during optimization.

\subsubsection{Attention Randomization Loss}
Beyond re-weighting attention toward or away from specific modalities, we consider an objective that directly disrupts the structure of attention itself. For each layer and head, we construct a randomized attention matrix $\tilde{A}_{l,h}$ with entries sampled uniformly within the observed range of $A_{l,h}$ and masked to preserve autoregressive structure:
\begin{equation}
\begin{aligned}
\tilde{A}_{l,h}
&= \mathrm{tril}\!\Big(
(\max(A_{l,h}) - \min(A_{l,h})) \\
&\qquad \cdot \mathrm{Uniform}(0,1) + \min(A_{l,h})
\Big).
\end{aligned}
\end{equation}
We then quantify the divergence using KL divergence:
\begin{equation}
\begin{aligned}
\mathcal{L}^{(\text{randatt})}(\delta)
&= \sum_{l=1}^{L} \sum_{h=1}^{H}
\mathrm{KL}\!\Big(
\mathrm{softmax}(A_{l,h})
\,\Big\|\, \\
&\qquad \mathrm{softmax}(\tilde{A}_{l,h})
\Big).
\end{aligned}
\end{equation}
Minimizing this loss pushes attention toward randomized configurations.

\subsubsection{Hidden-State Similarity Loss}
We define an adversarial objective that targets hidden representations across transformer layers. Prior work shows that steering hidden states can induce jailbreak behaviors in large language models~\cite{shu2025layerwiseperturbationssparseautoencoders}. Let $h_l(x)$ denote the hidden states produced by layer $l$. We define
\begin{equation}
\begin{aligned}
\mathcal{L}^{(\mathrm{hidden\text{-}cos})}(\tilde{x}_a, x_a)
&=
\frac{1}{L}\sum_{l=1}^{L}
\frac{1}{|h_l|}\sum_{i} \\
&\qquad \cos\Big(h_l^{(i)}(\tilde{x}_a),\, h_l^{(i)}(x_a)\Big).
\end{aligned}
\end{equation}
where $\cos(\mathbf{u}, \mathbf{v}) = \frac{\mathbf{u} \cdot \mathbf{v}}{\|\mathbf{u}\|_2 \|\mathbf{v}\|_2}$ denotes cosine similarity.
Minimizing this loss pushes attacked hidden representations away from their clean directions.

\subsubsection{Combined Loss}
We also consider a unified attack that jointly optimizes all proposed losses:
\begin{equation}
\begin{split}
\mathcal{L}^{(\mathrm{combined})}(\delta)
= &~
\mathcal{L}_{\text{negLM}} + \mathcal{L}^{(\cos)}(\tilde{x}_a, x_a) \\
&+ \mathcal{L}^{(\text{visionatt})}(\delta) + \mathcal{L}^{(\text{audioatt})}(\delta) \\
&+ \mathcal{L}^{(\text{randatt})}(\delta) \\
&+ \mathcal{L}^{(\mathrm{hidden\text{-}cos})}(\tilde{x}_a, x_a).
\end{split}
\end{equation}
This combined formulation aggregates complementary failure modes into a single adversarial signal. Detailed mathematical formulations for all loss functions are provided in Appendix~\ref{appendix:loss_formulation}, and the optimization algorithm is described in Appendix~\ref{appendix:algorithm}.

\section{Experimental Setup}
\label{sec:experiments}

\subsection{Models}
The primary model used for evaluation of all six attack objectives is VideoLLAMA2~\cite{damonlpsg2024videollama2}. We additionally use Qwen 2.5 Omni~\cite{Qwen2.5-Omni} and Qwen 3 Omni~\cite{Qwen3-Omni} to train attacks using $\mathcal{L}_{\text{negLM}}$ and $\mathcal{L}^{(\cos)}$ and to evaluate our attacks in the black-box setting. For Audio-Speech recongition evaluation, we use Whisper Large-v2 on LibriSpeech~\cite{panayotov2015librispeech}. Details about these models are provided in Appendix~\ref{appendix:models}.

\subsection{Datasets}
We evaluate on three benchmarks: AVQA~\cite{yang2022avqa} for multiple-choice question answering, AVSD~\cite{Alamri_2019_CVPR} for video-based textual summarization, and Music-AVQA~\cite{Li2022Learning} for short answers on musical performances. Further details are provided in Appendix~\ref{appendix:datasets}. Implementation details including hyperparameters and reproducibility settings are in Appendix~\ref{appendix:implementation}.

\subsection{Evaluation Metrics}
\label{sec:metrics}

\paragraph{Attack Success Rate}
We evaluate an untargeted, audio-only adversary using the Attack Success Rate (ASR), defined as the fraction of originally correct predictions that are flipped by the attack. Let $\mathcal{D}=\{(x_a^{(i)},x_v^{(i)},q^{(i)},a^{\star(i)})\}_{i=1}^N$ be the evaluation set. For each example $i$, let $\tilde{x}_a^{(i)} = x_a^{(i)} + \delta^{(i)}$ denote the perturbed audio. We define
\begin{equation}
\mathrm{CleanCorrect}^{(i)}
:= \mathbf{1}\!\left[\hat{a}(x_a^{(i)},x_v^{(i)},q^{(i)};\theta)=a^{\star(i)}\right],
\end{equation}
\begin{equation}
\begin{split}
\mathrm{AttackSuccess}^{(i)}
:= \mathbf{1}\Big[
\mathrm{CleanCorrect}^{(i)}=1 ~\wedge \\
\hat{a}(\tilde{x}_a^{(i)},x_v^{(i)},q^{(i)};\theta)\neq a^{\star(i)}
\Big].
\end{split}
\end{equation}
The Attack Success Rate is then
\begin{equation}\label{eq:asr}
\mathrm{ASR}
=
\frac{\sum_{i=1}^N \mathrm{AttackSuccess}^{(i)}}
     {\sum_{i=1}^N \mathrm{CleanCorrect}^{(i)}}.
\end{equation}

\paragraph{Other Metrics}
To quantify imperceptibility of audio perturbations, we use LPIPS computed on log-mel spectrograms and SI-SNR (Scale-Invariant Signal-to-Noise Ratio). For Whisper evaluation, we use word error rate (WER). For AVSD, evaluation is performed using LLM-as-a-judge. Instances where LLM judgment flips from "CORRECT" to "INCORRECT" are considered attack success. Further details about evaluation metrics are in Appendix~\ref{appendix:eval_metrics}, ~\ref{appendix:llm_judge}.

\section{Results and Analysis}
\label{sec:results}

We organize our findings into three categories: (1) attack effectiveness and transferability, (2) perceptual and optimization analysis, and (3) attention and hidden-state dynamics.

\subsection{Attack Effectiveness and Transferability}

\textsc{Encoder-Space Attacks Dominate.}
\textbf{Attacks that directly perturb audio encoder representations consistently outperform objectives operating at the output, attention, or hidden-state levels.} As shown in Table~\ref{tab:dataset_transfer}, the encoder-based cosine similarity loss $\mathcal{L}^{(\cos)}$ achieves an ASR of 89.12\% on AVQA, substantially exceeding attention-based and hidden-state objectives trained under identical conditions. This dominance persists when training on Music-AVQA (Table~\ref{tab:dataset_transfer_music_avqa}), where $\mathcal{L}^{(\cos)}$ attains 89.07\% ASR on the source domain. These results indicate that the audio encoder constitutes a critical vulnerability bottleneck: small directional shifts in encoder embedding space are sufficient to corrupt downstream cross-modal reasoning.

\begin{table}[ht]
\centering
\adjustbox{max width=\linewidth}{
\begin{tabular}{lccc}
\toprule
\textbf{Attack Objective} & \textbf{AVQA} & \textbf{Music-AVQA} & \textbf{AVSD}\\
\midrule
$\mathcal{L}_\text{negLM}$ & 10.27 & 10.74 & 50.43 \\
$\mathcal{L}^{(\text{cos})}$ & 89.12 & 12.92 & 57.03 \\
$\mathcal{L}^{(\text{visionatt})}$ & 18.72 & 7.94 & 49.34  \\
$\mathcal{L}^{(\text{audioatt})}$ & 56.21 & 9.81 & 43.85 \\
$\mathcal{L}^{(\text{randatt})}$ & 17.24 & 9.81 & 45.67 \\
$\mathcal{L}^{(\text{hidden-cos})}$ & 15.13 & 10.28 & 41.53  \\
$\mathcal{L}^{(\text{combined})}$ & \textbf{96.03} & \textbf{13.80} & \textbf{59.48} \\
\bottomrule
\end{tabular}
}
\caption{Attack success rate (\%) for different loss functions trained on VideoLLAMA2 for 150 epochs on AVQA. Evaluated on 2000 AVQA and Music-AVQA samples and full AVSD validation set.}
\label{tab:dataset_transfer}
\end{table}

\textsc{Limited Cross-Model Transferability.}
\textbf{Adversarial perturbations learned on one architecture fail to transfer effectively to other multimodal models.} Table~\ref{tab:blackbox} shows that attacks trained on VideoLLAMA2 achieve an ASR of 10.27\% on the source model but drop sharply when evaluated on Qwen 2.5 Omni (4.61\%) and Qwen 3 Omni (3.62\%). Similarly, perturbations optimized on Qwen models yield low ASR when applied to alternative architectures. This lack of transferability suggests that attacks exploit model-specific representational characteristics rather than universal multimodal vulnerabilities.

\begin{table}[ht]
\centering
\adjustbox{max width=\linewidth}{
\begin{tabular}{lccc}
\toprule
\textbf{Source Model} & \textbf{VideoLLAMA2} & \textbf{Qwen 2.5 Omni} & \textbf{Qwen 3 Omni}\\
\midrule
\textbf{VideoLLAMA2} & \textbf{10.27} & 4.61 & 3.62 \\
\textbf{Qwen 2.5 Omni} & 1.97 & \textbf{5.10}	& 1.12 \\
\textbf{Qwen 3 Omni} & 1.80 & 3.80 & \textbf{4.40} \\
\bottomrule
\end{tabular}
}
\caption{Black-box transfer results for $\mathcal{L}_\text{negLM}$. VL2 = VideoLLAMA2, Q2.5 = Qwen 2.5 Omni, Q3 = Qwen 3 Omni. Bold indicates source model (white-box).}
\label{tab:blackbox}
\end{table}

\textsc{Limited Cross-Domain Transfer.}
\textbf{Attacks trained on one domain exhibit limited transfer to domains with different acoustic characteristics.} From Table~\ref{tab:dataset_transfer}, attacks trained on AVQA transfer reasonably to AVSD (59.48\% for $\mathcal{L}^{(\text{combined})}$), but achieve only 13.80\% ASR on Music-AVQA, a dataset focused on musical performances. Similarly, Table~\ref{tab:dataset_transfer_music_avqa} shows that attacks trained on Music-AVQA achieve 89.07\% ASR on the source domain but only 2.30\% and 24.55\% on AVQA and AVSD respectively.

\begin{table}[ht]
\centering
\adjustbox{max width=\linewidth}{
\begin{tabular}{lccc}
\toprule
\textbf{Attack Objective} & \textbf{AVQA} & \textbf{Music-AVQA} & \textbf{AVSD}\\
\midrule
$\mathcal{L}_\text{negLM}$ & 2.00 & 16.60 & \textbf{30.88} \\
$\mathcal{L}^{(\text{cos})}$ & 2.30 & \textbf{89.07} & 24.55 \\
$\mathcal{L}^{(\text{hidden-cos})}$ & 2.20 & \textbf{55.32} & 21.93 \\
\bottomrule
\end{tabular}
}
\caption{ASR (\%) for attacks trained on Music-AVQA for 150 epochs.}
\label{tab:dataset_transfer_music_avqa}
\end{table}

\textsc{Encoder-Space Attacks Do Not Transfer Across Architectures.}
\textbf{Encoder-space adversarial perturbations are highly specific to the audio encoder on which they are optimized.} Table~\ref{tab:encoder_transferability} shows that attacks trained on the VideoLLAMA2 encoder reduce cosine similarity to 0.50 on the same encoder, but yield higher similarity when evaluated on Qwen 2.5 Omni (0.75), Qwen 3 Omni (0.70), or PANN (0.62) \cite{kong2020pannslargescalepretrainedaudio}. This pattern indicates that perturbations exploit encoder-specific feature geometries rather than inducing generic acoustic distortions.

\begin{table}[ht]
\centering
\adjustbox{max width=\linewidth}{
\begin{tabular}{lccc}
\toprule
\textbf{Encoder} & \textbf{VideoLLAMA2} & \textbf{Qwen 2.5 Omni} & \textbf{Qwen 3 Omni}\\
\midrule
\textbf{VideoLLAMA2} & \textbf{0.50} & 0.75 & 0.70 \\
\textbf{Qwen 2.5 Omni} & 0.77 & \textbf{0.74} & 0.75 \\
\textbf{Qwen 3 Omni} & 0.96 & 0.93 & \textbf{0.93} \\
\textbf{PANN} & 0.62 & 0.71 & 0.74 \\
\bottomrule
\end{tabular}
}
\caption{Cosine similarity across different audio encoders for $\mathcal{L}^{(\text{cos})}$ attacks. Lower similarity indicates stronger attack effect.}
\label{tab:encoder_transferability}
\end{table}

\subsection{Perceptual and Optimization Analysis}

\textsc{Effective Attacks Can Have Low Perceptual Distortion.}
\textbf{Highly effective attacks can be achieved with relatively small perceptual distortion to the original audio.} Table~\ref{tab:perceptibility} and Figure~\ref{plot:clean_vs_attacked_perturbation} (see Appendix~\ref{appendix:waveform} for detailed waveform analysis) show that encoder-space and attention-based attacks such as $\mathcal{L}^{(\cos)}$ and $\mathcal{L}^{(\text{audioatt})}$ achieve strong attack success while maintaining low LPIPS (0.08 and 0.06) and near-zero or positive SI-SNR ($-1.77$ and 0.33). In contrast, $\mathcal{L}_{\text{negLM}}$ incurs substantially higher distortion (LPIPS 0.22, SI-SNR $-11.48$). These results demonstrate that adversarial failures can arise from subtle, structured perturbations rather than large or perceptually dominant noise.

\begin{table}[ht]
\centering
\adjustbox{max width=\linewidth}{
\begin{tabular}{lcc}
\toprule
\textbf{Attack Objective} & \textbf{LPIPS} ($\downarrow$) & \textbf{SI-SNR (dB)} ($\uparrow$)\\
\midrule
$\mathcal{L}_\text{negLM}$ & 0.22 & $-11.48$\\
$\mathcal{L}^{(\text{cos})}$ & 0.08 & $-1.77$ \\
$\mathcal{L}^{(\text{visionatt})}$ & 0.07 & $-1.02$ \\
$\mathcal{L}^{(\text{audioatt})}$ & 0.06 & 0.33 \\
$\mathcal{L}^{(\text{randatt})}$ & 0.06 & 0.05\\
$\mathcal{L}^{(\text{hidden-cos})}$ & 0.06 & 0.56\\
$\mathcal{L}^{(\text{combined})}$ & 0.14 & $-6.23$\\
\bottomrule
\end{tabular}
}
\caption{LPIPS of log-mel spectrograms and SI-SNR for each loss trained on VideoLLAMA2. Lower LPIPS and higher SI-SNR indicate less perceptible perturbations.}
\label{tab:perceptibility}
\end{table}

\begin{figure}[ht]
    \centering
    \includegraphics[width=0.9\linewidth]{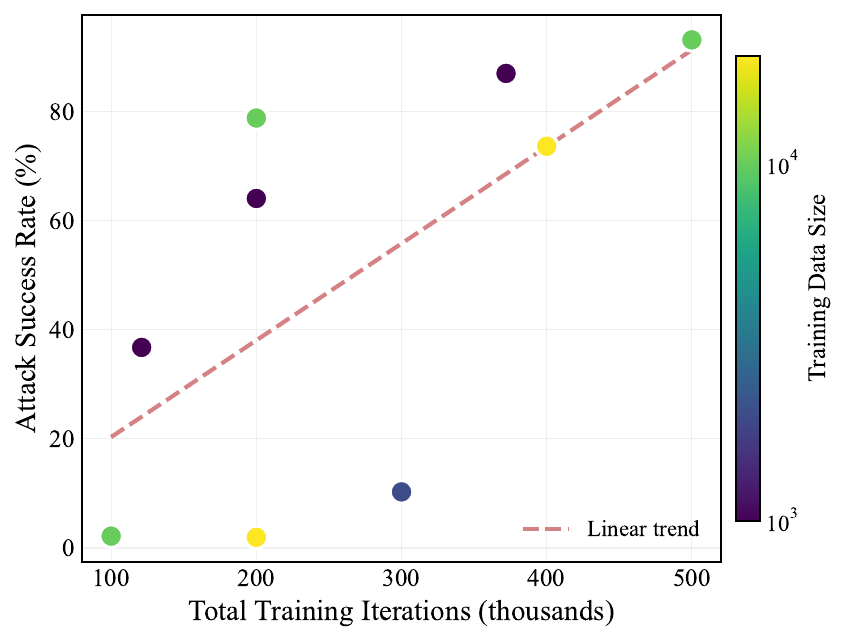}
    \caption{Relation between total training iterations and ASR for $\mathcal{L}_\text{negLM}$. Each point corresponds to an attack configuration, colored by training data size. Extended optimization on smaller datasets yields higher ASR.}
    \label{plot:iteration_vs_asr_loss1}
\end{figure}

\textsc{Extended Optimization Outperforms Larger Datasets.}
\textbf{Attack effectiveness is driven more by sustained optimization than by dataset size.} Figure~\ref{plot:iteration_vs_asr_loss1} shows that attacks trained on relatively small datasets but optimized for many iterations consistently achieve higher ASR than attacks trained on larger datasets for fewer epochs. Configurations with only a few thousand training samples but extended optimization reach ASR values exceeding 80\%, whereas attacks trained on tens of thousands of samples with limited iterations remain below 20\% ASR. This indicates that adversarial optimization benefits from repeatedly exploiting consistent model behaviors present in a small set of inputs. See App.~\ref{appendix:additional_results} for exact values.

\textsc{Non-Monotonic Attack Budget Effects.}
\textbf{Increasing the perturbation budget does not monotonically improve attack effectiveness.} For attack budget analysis, each attack is trained until it converges (conditions for convergence are in Appendix ~\ref{app:convergence}). Figure ~\ref{plot:budget_vs_asr} shows that ASR improves from 49.2\% at budget 0.3 to 73.51\% at budget 1.0, even though both settings converge in a similar number of epochs (213 vs.\ 209). In contrast, a mid-range budget of 0.7 underperforms substantially (51.89\% ASR) despite requiring more epochs to converge than 1.0 (265 vs.\ 209). This pattern suggests that the attack constraint interact with optimization dynamics in complex ways: some budgets admit highly effective perturbations that are also easier to optimize (budget 1.0), while others converge to weaker solutions even with extended training (budget 0.7). 

\begin{figure}[ht]
    \centering
    \includegraphics[width=0.9\linewidth]{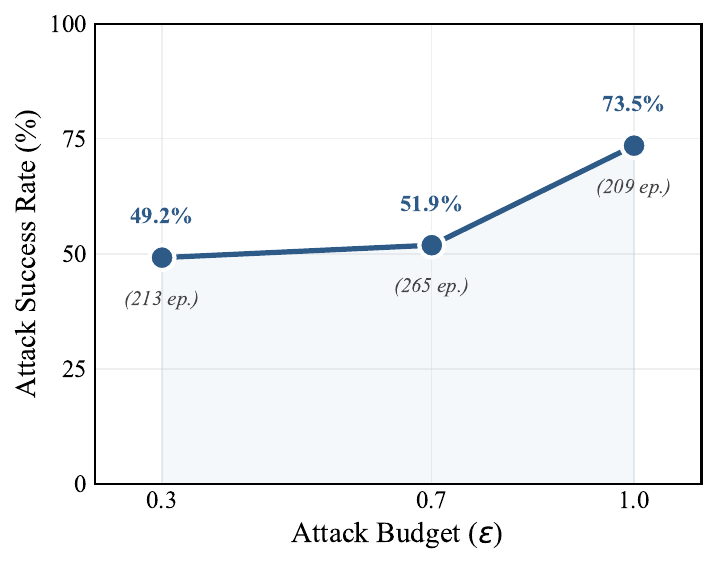}
    \caption{Attack budget vs ASR for $\mathcal{L}^{(\text{cos})}$ on VideoLLAMA2. Each attack was trained until convergence. }
    \label{plot:budget_vs_asr}
\end{figure}

\subsection{Attention, Hidden-State and Output Dynamics}

\textsc{Attack Success Arises from Distortions, Not Low Baseline.}
\textbf{High ASR is driven by attack-induced distortions rather than low baseline model performance.} Table~\ref{tab:accuracy} shows that the model maintains strong clean performance prior to attack (AVQA: 0.956, Music-AVQA: 0.807). Under adversarial perturbation, accuracy drops sharply for effective attacks: $\mathcal{L}^{(\cos)}$ reduces AVQA accuracy from 0.956 to 0.11, while the combined attack causes degradation to 0.039. In contrast, weaker objectives such as vision-attention manipulation produce only marginal accuracy changes (0.036 drop).

\textsc{Adversarial Responses Maintain High Confidence.}
\textbf{Adversarially induced responses are generated with confidence levels similar to clean inputs.} Table~\ref{tab:seq_confidence_full} shows that under the combined attack, the model's confidence on AVQA remains close between clean and adversarial inputs (0.93 vs.\ 0.85), despite a substantial ASR of 96.03\%. Several attacks exhibit small or even negative confidence differences, indicating that adversarial failures arise from internal misalignment rather than uncertainty. 

\textsc{Layer-Specific Vulnerabilities.}
\textbf{The effectiveness of attention and hidden-state attacks depends strongly on which layers are perturbed.} Table~\ref{tab:layer_subsets} shows that lower layers (1--10) contribute most to audio-driven attacks, with audio-attention and hidden-state losses achieving ASRs of 39.75\% and 32.85\%. Mid-level layers (11--18) are most influential for video-attention manipulation (27.92\% ASR). Higher layers (19--28) yield consistently low ASR across all attack types. When all layers are jointly optimized, attack success increases substantially, confirming that different layer subsets capture complementary vulnerabilities. See Appendix~\ref{appendix:layerwise} for layer-wise attention visualizations.

\begin{table}[ht]
\centering
\adjustbox{max width=\linewidth}{
\begin{tabular}{lccc}
\toprule
\textbf{Layers} & \textbf{Audio Attention} & \textbf{Video Attention} & \textbf{Hidden}\\
\midrule
1--10 & 39.75 & 2.25 & \textbf{32.85} \\
11--18 & 11.98 & \textbf{27.92} & 2.90 \\
19--28 & 2.04 & 2.20 & 2.62 \\
All & \textbf{56.21} & 18.72 & 15.13 \\
\bottomrule
\end{tabular}
}
\caption{ASR (\%) by layer subset. VL2 has 28 layers.}
\label{tab:layer_subsets}
\end{table}

\begin{figure}[ht]
    \centering
    \includegraphics[width=0.9\linewidth]{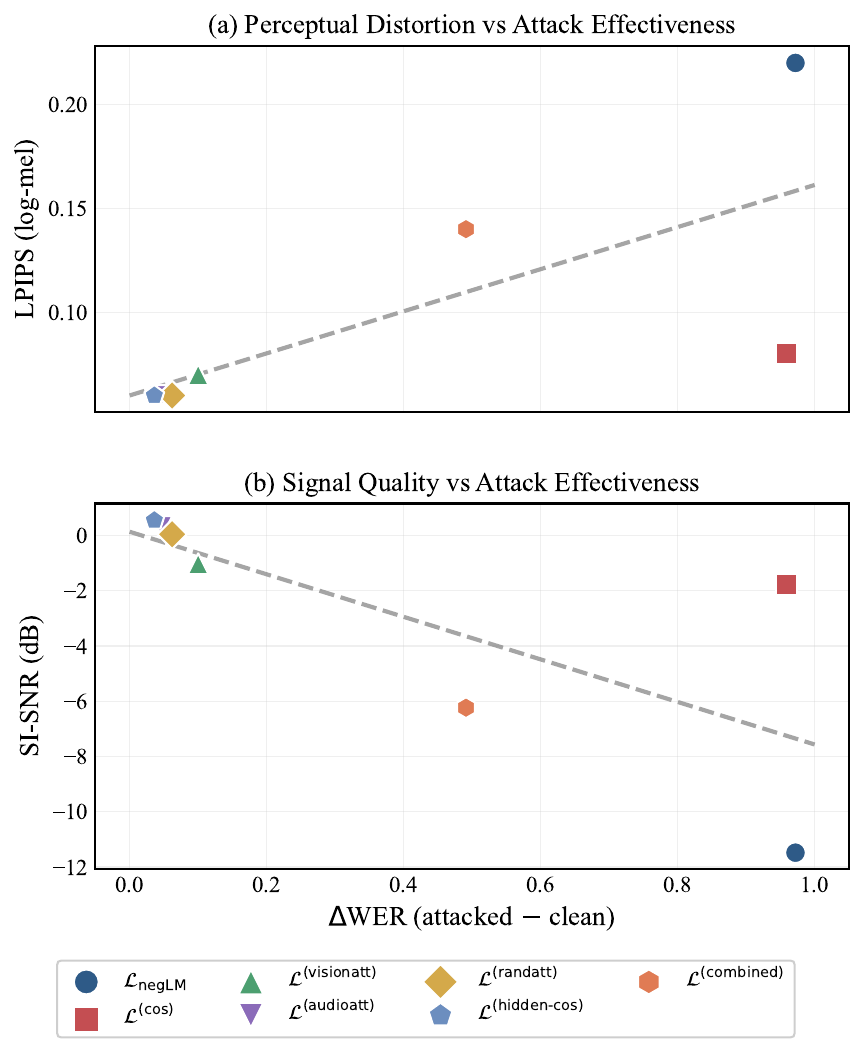}
    \caption{Relationship between attack effectiveness on Whisper and perceptual distortion. Top: WER difference vs LPIPS. Bottom: WER difference vs SI-SNR.}
    \label{plot:asr_vs_impercetibility}
\end{figure}

\textsc{Automatic Speech Recongition Systems Respond to Distortion Magnitude.}
\textbf{Speech recognition models are primarily sensitive to perturbation magnitude rather than specific attack objectives.} Figure~\ref{plot:asr_vs_impercetibility} and Table~\ref{tab:whisper_results} show that larger WER increases correlate strongly with higher perceptual distortion (LPIPS). Losses that induce substantial distortion, such as $\mathcal{L}_{\text{negLM}}$ and $\mathcal{L}^{(\cos)}$, achieve high ASR on Whisper (98.20\% and 97.38\%). In contrast, attention and hidden-state attacks produce minimal distortion and correspondingly low ASR. This indicates that automatic speech recognition systems are more sensitive to overall acoustic corruption than to structured adversarial manipulations.

\section{Conclusion}
\label{sec:conclusion}

We present \our, a systematic study of audio-only adversarial attacks on trimodal audio-video-language models. Our findings reveal that carefully crafted audio perturbations can reliably induce multimodal failures without modifying visual or textual inputs. Encoder-space attacks constitute a critical vulnerability bottleneck, achieving up to 96\% attack success rate. We show that these attacks exhibit limited transferability across model architectures and encoder designs, while speech recognition models respond primarily to distortion magnitude rather than structured adversarial objectives. Our analysis of layer-specific vulnerabilities, optimization dynamics, and confidence patterns provides practical insights for both attack construction and defensive design. Future work should explore adaptive defenses that monitor cross-modal consistency and investigate attack persistence under real-world conditions such as over-the-air propagation and environmental interference.

\section*{Acknowledgments}
We acknowledge Advanced Research Computing at Virginia Tech for providing computational resources and technical support that have contributed to the results reported within this paper. We also thank all reviewers for their comments which helped improve the paper.

\section*{Limitations}
Our study has several limitations that warrant discussion. \textbf{First,} we focus exclusively on white-box attacks where the adversary has full access to model gradients and internal representations. Real-world adversaries may have limited or no access to model internals, and developing effective black-box attacks remains an open challenge. While we evaluate black-box transferability across models, the low transfer rates suggest that practical deployment would require model-specific optimization. \textbf{Second,} all experiments apply perturbations digitally to audio waveforms. Physical-world attacks face additional challenges including environmental noise, room reverberation, speaker-microphone frequency response variations, and lossy audio compression. The extent to which our attacks remain effective under such conditions requires further investigation. \textbf{Third,} our evaluation is limited to three multimodal models and three benchmarks. Generalization to other architectures, particularly those with fundamentally different audio encoding strategies or cross-modal fusion mechanisms, cannot be assumed without additional experiments. \textbf{Fourth,} we do not evaluate potential defenses against our attacks. Promising directions include input preprocessing, adversarial training, and cross-modal consistency checking, but their efficacy and computational costs remain unexplored.

\section*{Ethics Statement}
This work studies adversarial vulnerabilities in multimodal models to inform the development of more robust systems. We do not provide tools or code that could be directly used for malicious attacks. Our evaluation uses publicly available models and datasets, requiring no collection of personal data or human annotations. All models are evaluated in accordance with their respective licensing agreements. We believe that understanding these vulnerabilities is essential for deploying multimodal systems safely, and that responsible disclosure of attack surfaces enables the research community to develop appropriate defenses.

\bibliography{custom}

@article{liu2025survey,
  title={A Survey of Attacks on Large Vision--Language Models: Resources, Advances, and Future Trends},
  author={Liu, Daizong and Yang, Mingyu and Qu, Xiaoye and Zhou, Pan and Cheng, Yu and Hu, Wei},
  journal={IEEE Transactions on Neural Networks and Learning Systems},
  year={2025},
  publisher={IEEE}
}

@misc{shu2025layerwiseperturbationssparseautoencoders,
      title={Layer-Wise Perturbations via Sparse Autoencoders for Adversarial Text Generation}, 
      author={Huizhen Shu and Xuying Li and Qirui Wang and Yuji Kosuga and Mengqiu Tian and Zhuo Li},
      year={2025},
      eprint={2508.10404},
      archivePrefix={arXiv},
      primaryClass={cs.CL},
      url={https://arxiv.org/abs/2508.10404}, 
}

@inproceedings{panayotov2015librispeech,
  title={Librispeech: an ASR corpus based on public domain audio books},
  author={Panayotov, Vassil and Chen, Guoguo and Povey, Daniel and Khudanpur, Sanjeev},
  booktitle={Acoustics, Speech and Signal Processing (ICASSP), 2015 IEEE International Conference on},
  pages={5206--5210},
  year={2015},
  organization={IEEE}
}

@inproceedings{Li2022Learning,
  title={Learning to answer questions in dynamic audio-visual scenarios},
  author={Li, Guangyao and Wei, Yake and Tian, Yapeng and Xu, Chenliang and Wen, Ji-Rong and Hu, Di},
  booktitle={Proceedings of the IEEE/CVF conference on computer vision and pattern recognition},
  pages={19108--19118},
  year={2022}
}

@misc{geisler2025attackinglargelanguagemodels,
      title={Attacking Large Language Models with Projected Gradient Descent}, 
      author={Simon Geisler and Tom Wollschläger and M. H. I. Abdalla and Johannes Gasteiger and Stephan Günnemann},
      year={2025},
      eprint={2402.09154},
      archivePrefix={arXiv},
      primaryClass={cs.LG},
      url={https://arxiv.org/abs/2402.09154}, 
}

@misc{wang2025attentionvisionlanguagemodel,
      title={Attention! Your Vision Language Model Could Be Maliciously Manipulated}, 
      author={Xiaosen Wang and Shaokang Wang and Zhijin Ge and Yuyang Luo and Shudong Zhang},
      year={2025},
      eprint={2505.19911},
      archivePrefix={arXiv},
      primaryClass={cs.CV},
      url={https://arxiv.org/abs/2505.19911}, 
}

@misc{wang2024attngcgenhancingjailbreakingattacks,
      title={AttnGCG: Enhancing Jailbreaking Attacks on LLMs with Attention Manipulation}, 
      author={Zijun Wang and Haoqin Tu and Jieru Mei and Bingchen Zhao and Yisen Wang and Cihang Xie},
      year={2024},
      eprint={2410.09040},
      archivePrefix={arXiv},
      primaryClass={cs.CL},
      url={https://arxiv.org/abs/2410.09040}, 
}

@misc{yi2024jailbreakattacksdefenseslarge,
      title={Jailbreak Attacks and Defenses Against Large Language Models: A Survey}, 
      author={Sibo Yi and Yule Liu and Zhen Sun and Tianshuo Cong and Xinlei He and Jiaxing Song and Ke Xu and Qi Li},
      year={2024},
      eprint={2407.04295},
      archivePrefix={arXiv},
      primaryClass={cs.CR},
      url={https://arxiv.org/abs/2407.04295}, 
}

@misc{wang2024whiteboxmultimodaljailbreakslarge,
      title={White-box Multimodal Jailbreaks Against Large Vision-Language Models}, 
      author={Ruofan Wang and Xingjun Ma and Hanxu Zhou and Chuanjun Ji and Guangnan Ye and Yu-Gang Jiang},
      year={2024},
      eprint={2405.17894},
      archivePrefix={arXiv},
      primaryClass={cs.CV},
      url={https://arxiv.org/abs/2405.17894}, 
}

@misc{bagdasaryan2023abusingimagessoundsindirect,
      title={Abusing Images and Sounds for Indirect Instruction Injection in Multi-Modal LLMs}, 
      author={Eugene Bagdasaryan and Tsung-Yin Hsieh and Ben Nassi and Vitaly Shmatikov},
      year={2023},
      eprint={2307.10490},
      archivePrefix={arXiv},
      primaryClass={cs.CR},
      url={https://arxiv.org/abs/2307.10490}, 
}

@inproceedings{walmer2022dual,
  title={Dual-key multimodal backdoors for visual question answering},
  author={Walmer, Matthew and Sikka, Karan and Sur, Indranil and Shrivastava, Abhinav and Jha, Susmit},
  booktitle={Proceedings of the IEEE/CVF Conference on computer vision and pattern recognition},
  pages={15375--15385},
  year={2022}
}

@article{xu2024shadowcast,
  title={Shadowcast: Stealthy data poisoning attacks against vision-language models},
  author={Xu, Yuancheng and Yao, Jiarui and Shu, Manli and Sun, Yanchao and Wu, Zichu and Yu, Ning and Goldstein, Tom and Huang, Furong},
  journal={Advances in Neural Information Processing Systems},
  volume={37},
  pages={57733--57764},
  year={2024}
}

@misc{chary2025spectrogrampatchcodec2d,
      title={Spectrogram Patch Codec: A 2D Block-Quantized VQ-VAE and HiFi-GAN for Neural Speech Coding}, 
      author={Luis Felipe Chary and Miguel Arjona Ramirez},
      year={2025},
      eprint={2509.02244},
      archivePrefix={arXiv},
      primaryClass={cs.SD},
      url={https://arxiv.org/abs/2509.02244}, 
}

@misc{kong2020pannslargescalepretrainedaudio,
      title={PANNs: Large-Scale Pretrained Audio Neural Networks for Audio Pattern Recognition}, 
      author={Qiuqiang Kong and Yin Cao and Turab Iqbal and Yuxuan Wang and Wenwu Wang and Mark D. Plumbley},
      year={2020},
      eprint={1912.10211},
      archivePrefix={arXiv},
      primaryClass={cs.SD},
      url={https://arxiv.org/abs/1912.10211}, 
}

@misc{mustakim2025watchlistenunderstandmislead,
      title={Watch, Listen, Understand, Mislead: Tri-modal Adversarial Attacks on Short Videos for Content Appropriateness Evaluation}, 
      author={Sahid Hossain Mustakim and S M Jishanul Islam and Ummay Maria Muna and Montasir Chowdhury and Mohammed Jawwadul Islam and Sadia Ahmmed and Tashfia Sikder and Syed Tasdid Azam Dhrubo and Swakkhar Shatabda},
      year={2025},
      eprint={2507.11968},
      archivePrefix={arXiv},
      primaryClass={cs.CV},
      url={https://arxiv.org/abs/2507.11968}, 
}

@misc{goodfellow2015explainingharnessingadversarialexamples,
      title={Explaining and Harnessing Adversarial Examples}, 
      author={Ian J. Goodfellow and Jonathon Shlens and Christian Szegedy},
      year={2015},
      eprint={1412.6572},
      archivePrefix={arXiv},
      primaryClass={stat.ML},
      url={https://arxiv.org/abs/1412.6572}, 
}

@inproceedings{carlini2017towards,
  title={Towards evaluating the robustness of neural networks},
  author={Carlini, Nicholas and Wagner, David},
  booktitle={2017 ieee symposium on security and privacy (sp)},
  pages={39--57},
  year={2017},
  organization={Ieee}
}

@inproceedings{geisler2024attacking,
  title={Attacking Large Language Models with Projected Gradient Descent},
  author={Geisler, Simon and Wollschl{\"a}ger, Tom and Abdalla, MHI and Gasteiger, Johannes and G{\"u}nnemann, Stephan},
  booktitle={ICML 2024 Next Generation of AI Safety Workshop},
  year={2024}
}

@inproceedings{zhang2025qava,
  title={QAVA: Query-Agnostic Visual Attack to Large Vision-Language Models},
  author={Zhang, Yudong and Xie, Ruobing and Chen, Jiansheng and Sun, Xingwu and Kang, Zhanhui and Wang, Yu},
  booktitle={Proceedings of the 2025 Conference of the Nations of the Americas Chapter of the Association for Computational Linguistics: Human Language Technologies (Volume 1: Long Papers)},
  pages={10205--10218},
  year={2025}
}

@inproceedings{zhang2025anyattack,
  title={AnyAttack: Towards Large-scale Self-supervised Adversarial Attacks on Vision-language Models},
  author={Zhang, Jiaming and Ye, Junhong and Ma, Xingjun and Li, Yige and Yang, Yunfan and Chen, Yunhao and Sang, Jitao and Yeung, Dit-Yan},
  booktitle={Proceedings of the Computer Vision and Pattern Recognition Conference},
  pages={19900--19909},
  year={2025}
}

@inproceedings{hao2025exploring,
  title={Exploring Visual Vulnerabilities via Multi-Loss Adversarial Search for Jailbreaking Vision-Language Models},
  author={Hao, Shuyang and Hooi, Bryan and Liu, Jun and Chang, Kai-Wei and Huang, Zi and Cai, Yujun},
  booktitle={Proceedings of the Computer Vision and Pattern Recognition Conference},
  pages={19890--19899},
  year={2025}
}

@article{mei2025veattack,
  title={VEAttack: Downstream-agnostic Vision Encoder Attack against Large Vision Language Models},
  author={Mei, Hefei and Wang, Zirui and You, Shen and Dong, Minjing and Xu, Chang},
  journal={arXiv preprint arXiv:2505.17440},
  year={2025}
}

@article{sadasivan2025attacker,
  title={Attacker's Noise Can Manipulate Your Audio-based LLM in the Real World},
  author={Sadasivan, Vinu Sankar and Feizi, Soheil and Mathews, Rajiv and Wang, Lun},
  journal={arXiv preprint arXiv:2507.06256},
  year={2025}
}

@article{djanibekov2025spirit,
  title={SPIRIT: Patching Speech Language Models against Jailbreak Attacks},
  author={Djanibekov, Amirbek and Mukhituly, Nurdaulet and Inui, Kentaro and Aldarmaki, Hanan and Lukas, Nils},
  journal={arXiv preprint arXiv:2505.13541},
  year={2025}
}

@article{ma2025universal,
  title={Universal Acoustic Adversarial Attacks for Flexible Control of Speech-LLMs},
  author={Ma, Rao and Qian, Mengjie and Raina, Vyas and Gales, Mark and Knill, Kate},
  journal={arXiv preprint arXiv:2505.14286},
  year={2025}
}

@article{liu2025exploiting,
  title={Exploiting Vulnerabilities in Speech Translation Systems through Targeted Adversarial Attacks},
  author={Liu, Chang and Wu, Haolin and Yang, Xi and Zhang, Kui and Wu, Cong and Zhang, Weiming and Yu, Nenghai and Zhang, Tianwei and Guo, Qing and Zhang, Jie},
  journal={arXiv preprint arXiv:2503.00957},
  year={2025}
}

@inproceedings{raina2024muting,
  title={Muting Whisper: A Universal Acoustic Adversarial Attack on Speech Foundation Models},
  author={Raina, Vyas and Ma, Rao and McGhee, Charles and Knill, Kate and Gales, Mark},
  booktitle={Proceedings of the 2024 Conference on Empirical Methods in Natural Language Processing},
  pages={7549--7565},
  year={2024}
}

@inproceedings{kangadvwave,
  title={AdvWave: Stealthy Adversarial Jailbreak Attack against Large Audio-Language Models},
  author={Kang, Mintong and Xu, Chejian and Li, Bo},
  booktitle={The Thirteenth International Conference on Learning Representations},
  year={2025}
}

@article{szegedy2013intriguing,
  title={Intriguing properties of neural networks},
  author={Szegedy, Christian and Zaremba, Wojciech and Sutskever, Ilya and Bruna, Joan and Erhan, Dumitru and Goodfellow, Ian and Fergus, Rob},
  journal={arXiv preprint arXiv:1312.6199},
  year={2013}
}

@inproceedings{nguyen2015deep,
  title={Deep neural networks are easily fooled: High confidence predictions for unrecognizable images},
  author={Nguyen, Anh and Yosinski, Jason and Clune, Jeff},
  booktitle={Proceedings of the IEEE conference on computer vision and pattern recognition},
  pages={427--436},
  year={2015}
}

@inproceedings{papernot2016limitations,
  title={The limitations of deep learning in adversarial settings},
  author={Papernot, Nicolas and McDaniel, Patrick and Jha, Somesh and Fredrikson, Matt and Celik, Z Berkay and Swami, Ananthram},
  booktitle={2016 IEEE European symposium on security and privacy (EuroS\&P)},
  pages={372--387},
  year={2016},
  organization={IEEE}
}

@inproceedings{moosavi2016deepfool,
  title={Deepfool: a simple and accurate method to fool deep neural networks},
  author={Moosavi-Dezfooli, Seyed-Mohsen and Fawzi, Alhussein and Frossard, Pascal},
  booktitle={Proceedings of the IEEE conference on computer vision and pattern recognition},
  pages={2574--2582},
  year={2016}
}

@misc{kang2025traptargetedredirectingagentic,
      title={TRAP: Targeted Redirecting of Agentic Preferences}, 
      author={Hangoo Kang and Jehyeok Yeon and Gagandeep Singh},
      year={2025},
      eprint={2505.23518},
      archivePrefix={arXiv},
      primaryClass={cs.AI},
      url={https://arxiv.org/abs/2505.23518}, 
}

@article{choi2024ghost,
  title={Ghost in the radio: An audio adversarial attack using environmental noise through radio},
  author={Choi, Hyeongjun and Jung, Ji Hyuk and Yoon, Ji Won},
  journal={IEEE Access},
  year={2024},
  publisher={IEEE}
}

@article{damonlpsg2024videollama2,
  title={VideoLLaMA 2: Advancing Spatial-Temporal Modeling and Audio Understanding in Video-LLMs},
  author={Cheng, Zesen and Leng, Sicong and Zhang, Hang and Xin, Yifei and Li, Xin and Chen, Guanzheng and Zhu, Yongxin and Zhang, Wenqi and Luo, Ziyang and Zhao, Deli and Bing, Lidong},
  journal={arXiv preprint arXiv:2406.07476},
  year={2024},
  url = {https://arxiv.org/abs/2406.07476}
}

@inproceedings{yang2022avqa,
  title={Avqa: A dataset for audio-visual question answering on videos},
  author={Yang, Pinci and Wang, Xin and Duan, Xuguang and Chen, Hong and Hou, Runze and Jin, Cong and Zhu, Wenwu},
  booktitle={Proceedings of the 30th ACM international conference on multimedia},
  pages={3480--3491},
  year={2022}
}

@InProceedings{Alamri_2019_CVPR,
author = {Alamri, Huda and Cartillier, Vincent and Das, Abhishek and Wang, Jue and Cherian, Anoop and Essa, Irfan and Batra, Dhruv and Marks, Tim K. and Hori, Chiori and Anderson, Peter and Lee, Stefan and Parikh, Devi},
title = {Audio Visual Scene-Aware Dialog},
booktitle = {Proceedings of the IEEE/CVF Conference on Computer Vision and Pattern Recognition (CVPR)},
month = {June},
year = {2019}
}

@misc{Qwen2.5-Omni,
      title={Qwen2.5-Omni Technical Report}, 
      author={Jin Xu and Zhifang Guo and Jinzheng He and Hangrui Hu and Ting He and Shuai Bai and Keqin Chen and Jialin Wang and Yang Fan and Kai Dang and Bin Zhang and Xiong Wang and Yunfei Chu and Junyang Lin},
      year={2025},
      eprint={2503.20215},
      archivePrefix={arXiv},
      primaryClass={cs.CL},
      url={https://arxiv.org/abs/2503.20215}, 
}

@misc{goodfellow2014explaining,
      title={Explaining and Harnessing Adversarial Examples}, 
      author={Ian J. Goodfellow and Jonathon Shlens and Christian Szegedy},
      year={2015},
      eprint={1412.6572},
      archivePrefix={arXiv},
      primaryClass={stat.ML},
      url={https://arxiv.org/abs/1412.6572}, 
}

@article{Qwen3-Omni,
  title={Qwen3-Omni Technical Report},
  author={Jin Xu and Zhifang Guo and Hangrui Hu and Yunfei Chu and Xiong Wang and Jinzheng He and Yuxuan Wang and Xian Shi and Ting He and Xinfa Zhu and Yuanjun Lv and Yongqi Wang and Dake Guo and He Wang and Linhan Ma and Pei Zhang and Xinyu Zhang and Hongkun Hao and Zishan Guo and Baosong Yang and Bin Zhang and Ziyang Ma and Xipin Wei and Shuai Bai and Keqin Chen and Xuejing Liu and Peng Wang and Mingkun Yang and Dayiheng Liu and Xingzhang Ren and Bo Zheng and Rui Men and Fan Zhou and Bowen Yu and Jianxin Yang and Le Yu and Jingren Zhou and Junyang Lin},
  journal={arXiv preprint arXiv:2509.17765},
  year={2025}
}

@article{shayegani2023jailbreak,
  title={Jailbreak in pieces: Compositional adversarial attacks on multi-modal language models},
  author={Shayegani, Erfan and Dong, Yue and Abu-Ghazaleh, Nael},
  journal={arXiv preprint arXiv:2307.14539},
  year={2023}
}

@article{jiang2025survey,
  title={Survey of adversarial robustness in multimodal large language models},
  author={Jiang, Chengze and Wang, Zhuangzhuang and Dong, Minjing and Gui, Jie},
  journal={arXiv preprint arXiv:2503.13962},
  year={2025}
}

@article{lu2025adversarial,
  title={Adversarial training for multimodal large language models against jailbreak attacks},
  author={Lu, Liming and Pang, Shuchao and Liang, Siyuan and Zhu, Haotian and Zeng, Xiyu and Liu, Aishan and Liu, Yunhuai and Zhou, Yongbin},
  journal={arXiv preprint arXiv:2503.04833},
  year={2025}
}

@article{liu2024jailbreak,
  title={Jailbreak attacks and defenses against multimodal generative models: A survey},
  author={Liu, Xuannan and Cui, Xing and Li, Peipei and Li, Zekun and Huang, Huaibo and Xia, Shuhan and Zhang, Miaoxuan and Zou, Yueying and He, Ran},
  journal={arXiv preprint arXiv:2411.09259},
  year={2024}
}

@article{zhang2025modality,
  title={Modality-Specific Interactive Attack for Vision-Language Pre-Training Models},
  author={Zhang, Haiqi and Tang, Hao and Sun, Yanpeng and He, Shengfeng and Li, Zechao},
  journal={IEEE Transactions on Information Forensics and Security},
  year={2025},
  publisher={IEEE}
}

@inproceedings{zhang2022towards,
  title={Towards adversarial attack on vision-language pre-training models},
  author={Zhang, Jiaming and Yi, Qi and Sang, Jitao},
  booktitle={Proceedings of the 30th ACM International Conference on Multimedia},
  pages={5005--5013},
  year={2022}
}

@article{liu2023privacy,
  title={Privacy-preserving and Privacy-attacking Approaches for Speech and Audio--A Survey},
  author={Liu, Yuchen and Kapadia, Apu and Williamson, Donald},
  journal={arXiv preprint arXiv:2309.15087},
  year={2023}
}

@inproceedings{wang2025manipulating,
  title={Manipulating multimodal agents via cross-modal prompt injection},
  author={Wang, Le and Ying, Zonghao and Zhang, Tianyuan and Liang, Siyuan and Hu, Shengshan and Zhang, Mingchuan and Liu, Aishan and Liu, Xianglong},
  booktitle={Proceedings of the 33rd ACM International Conference on Multimedia},
  pages={10955--10964},
  year={2025}
}

@inproceedings{tian2021can,
  title={Can audio-visual integration strengthen robustness under multimodal attacks?},
  author={Tian, Yapeng and Xu, Chenliang},
  booktitle={Proceedings of the IEEE/CVF conference on computer vision and pattern recognition},
  pages={5601--5611},
  year={2021}
}

@inproceedings{han2024backdooring,
  title={Backdooring multimodal learning},
  author={Han, Xingshuo and Wu, Yutong and Zhang, Qingjie and Zhou, Yuan and Xu, Yuan and Qiu, Han and Xu, Guowen and Zhang, Tianwei},
  booktitle={2024 IEEE Symposium on Security and Privacy (SP)},
  pages={3385--3403},
  year={2024},
  organization={IEEE}
}

@inproceedings{liang2024badclip,
  title={Badclip: Dual-embedding guided backdoor attack on multimodal contrastive learning},
  author={Liang, Siyuan and Zhu, Mingli and Liu, Aishan and Wu, Baoyuan and Cao, Xiaochun and Chang, Ee-Chien},
  booktitle={Proceedings of the IEEE/CVF conference on computer vision and pattern recognition},
  pages={24645--24654},
  year={2024}
}

@inproceedings{yu2025backdoormbti,
  title={BackdoorMBTI: A Backdoor Learning Multimodal Benchmark Tool Kit for Backdoor Defense Evaluation},
  author={Yu, Haiyang and Xie, Tian and Gui, Jiaping and Wang, Pengyang and Cheng, Pengzhou and Yi, Ping and Wu, Yue},
  booktitle={Proceedings of the 31st ACM SIGKDD Conference on Knowledge Discovery and Data Mining V. 1},
  pages={2791--2802},
  year={2025}
}

@article{liang2025vl,
  title={Vl-trojan: Multimodal instruction backdoor attacks against autoregressive visual language models},
  author={Liang, Jiawei and Liang, Siyuan and Liu, Aishan and Cao, Xiaochun},
  journal={International Journal of Computer Vision},
  pages={1--20},
  year={2025},
  publisher={Springer}
}

\appendix
\clearpage

\addtocontents{toc}{\protect\setcounter{tocdepth}{2}}

\renewcommand{\contentsname}{Contents of the Appendix}

\tableofcontents

\appendix

\section{Experimental Setup Details}
\label{appendix:setup}

\subsection{Model Architectures}
\label{appendix:models}

We evaluate on three multimodal models and two audio-only systems. Table~\ref{tab:model_summary} summarizes the key architectural differences.

\begin{table}[ht]
\centering
\adjustbox{max width=\linewidth}{
\begin{tabular}{lccl}
\toprule
\textbf{Model} & \textbf{Layers} & \textbf{Audio Encoder} & \textbf{LM Backbone} \\
\midrule
VideoLLAMA2 & 28 & BEATs & Qwen2-7B-Instruct \\
Qwen 2.5 Omni & 28 & Whisper-style & Qwen2.5 \\
Qwen 3 Omni & 28 & Whisper-style & Qwen3 \\
Whisper Large-v2 & 32 & Whisper encoder & Whisper decoder \\
PANNs & 14 & CNN & N/A \\
\bottomrule
\end{tabular}
}
\caption{Summary of model architectures used in our experiments.}
\label{tab:model_summary}
\end{table}

\paragraph{VideoLLAMA2.}
VideoLLAMA2~\cite{damonlpsg2024videollama2} serves as our primary evaluation model. The architecture uses a CLIP-based visual encoder and a separate audio encoder that extracts 128-dimensional mel-filterbank features. Audio embeddings are projected into the language model space and concatenated with visual tokens before processing by a Qwen2-7B-Instruct backbone. The model has 28 transformer layers, which we partition into early (1--10), middle (11--18), and late (19--28) subsets for layer-wise analysis.

\paragraph{Qwen 2.5 Omni and Qwen 3 Omni.}
Qwen 2.5 Omni~\cite{Qwen2.5-Omni} and Qwen 3 Omni~\cite{Qwen3-Omni} use a Thinker-Talker architecture where the Thinker processes multimodal inputs and the Talker generates text or speech outputs. Both models use a Whisper-style audio encoder with different tokenization than VideoLLAMA2. We use these models to evaluate whether attacks transfer across different audio processing pipelines. Qwen 3 Omni is a more recent release with updated training, providing a test case for transfer to newer model versions.

\paragraph{Whisper Large-v2.}
Whisper is an encoder-decoder speech recognition model trained on 680,000 hours of web audio. The encoder processes 80-dimensional log-mel spectrograms; the decoder generates transcriptions autoregressively. We use Whisper to evaluate black-box transfer of attacks to speech-only systems outside the trimodal setting.

\paragraph{PANNs.}
Pretrained Audio Neural Networks (PANNs)~\cite{kong2020pannslargescalepretrainedaudio} are CNNs trained on AudioSet for general audio tagging. We use PANNs as an external encoder to test whether $\mathcal{L}^{(\text{cos})}$ attacks transfer across encoder architectures or exploit VideoLLAMA2-specific features.

\section{Dataset Descriptions}
\label{appendix:datasets}

This section provides detailed descriptions of the datasets used for training and evaluating our adversarial attacks. We select datasets that span different audio-visual reasoning tasks to comprehensively assess attack effectiveness and transferability.

\paragraph{AVQA (Audio-Visual Question Answering).}
AVQA~\cite{yang2022avqa} is a large-scale audio-visual question answering dataset introduced at ACM MM 2022. The dataset contains approximately 57,015 videos sourced from VGG-Sound, depicting real audio-visual scenes with natural sounds and objects. Each video is associated with one or more question-answer pairs, totaling 57,335 QA instances. The questions are designed such that answering them often requires integrating both audio and visual cues; many questions cannot be answered correctly using a single modality alone. This cross-modal dependency makes AVQA particularly suitable for evaluating whether audio-only perturbations can disrupt multimodal reasoning even when visual information remains unperturbed. We use the original train/validation/test splits provided by the dataset authors. In our experiments, AVQA serves as the primary benchmark for both attack training and evaluation.

\paragraph{Music-AVQA.}
Music-AVQA~\cite{Li2022Learning} is a short-answer audio-visual question answering dataset focused on musical performance videos. The dataset requires fine-grained spatio-temporal reasoning across modalities, with questions that probe understanding of musical instruments, performer actions, and acoustic properties. Videos feature diverse musical performances including solo instruments, ensembles, and various musical genres. The acoustic characteristics of musical content differ substantially from the speech and environmental sounds that dominate AVQA, providing a challenging test case for cross-domain transfer. We use Music-AVQA to evaluate whether attacks trained on general audio-visual scenarios generalize to structured music-centric scenes with distinct spectral properties.

\paragraph{AVSD (Audio-Visual Scene-Aware Dialog).}
AVSD~\cite{Alamri_2019_CVPR} is a dialog-based dataset from DSTC7/DSTC8 where models generate free-form, natural language responses grounded in video content, audio information, and dialog history. The dataset is built over more than 11,000 videos from the Charades dataset, which depicts people performing everyday activities in home environments. Each sample includes a 10-round dialog history plus a final summary describing the video content. Unlike AVQA and Music-AVQA which use short-answer formats, AVSD requires generating extended natural language descriptions, providing a complementary evaluation of attack effectiveness on generative tasks. We use AVSD to assess cross-dataset transferability within the same domain, particularly for evaluating whether attacks transfer across different output modalities (multiple-choice versus open-ended generation).

\paragraph{LibriSpeech.}
LibriSpeech~\cite{panayotov2015librispeech} is a large-scale corpus of read English speech derived from audiobooks in the LibriVox project. The dataset contains approximately 1,000 hours of speech at 16kHz sampling rate, with high-quality transcriptions suitable for training and evaluating automatic speech recognition systems. The dataset is partitioned into training, development, and test sets with varying levels of difficulty based on speaker characteristics and recording conditions. In our experiments, LibriSpeech is used solely for evaluating black-box transfer of adversarial perturbations to Whisper, enabling controlled analysis of attack effectiveness in a speech-only setting without the confounding factors of visual input or multimodal reasoning.

\section{Mathematical Formulation of Attack Objectives}
\label{appendix:loss_formulation}

This section provides detailed mathematical formulations for each adversarial objective used in our attack framework. We describe the theoretical motivation, precise mathematical definitions, and implementation considerations for each loss function.

\subsection{Perturbation Model and Constraint Set}

Let $x_a \in \mathbb{R}^T$ denote the clean audio waveform of length $T$ samples at sampling rate $f_s = 16\text{kHz}$. The adversary constructs a perturbation $\delta \in \mathbb{R}^{T_\delta}$ of fixed length $T_\delta = 10 \cdot f_s = 160{,}000$ samples, corresponding to a 10-second audio segment. For audio inputs longer than $T_\delta$, the perturbation is repeated cyclically:
\begin{equation}
\delta_{\text{extended}}[t] = \delta[t \mod T_\delta], \quad t = 0, 1, \ldots, T-1.
\end{equation}
The perturbed audio is then computed as the element-wise sum:
\begin{equation}
\tilde{x}_a = x_a + \cdot \delta_{\text{extended}}[:T],
\end{equation}
The notation $[:T]$ denotes truncation to the first $T$ elements.

To ensure bounded perturbation magnitude, we constrain $\delta$ to lie within an $\ell_\infty$ ball:
\begin{equation}
\mathcal{C}_\varepsilon = \{\delta \in \mathbb{R}^{T_\delta} : \|\delta\|_\infty \leq \varepsilon\},
\end{equation}
where $\varepsilon > 0$ is the attack budget. After each gradient update, the perturbation is projected back onto $\mathcal{C}_\varepsilon$ via element-wise clipping:
\begin{equation}
\Pi_{\mathcal{C}_\varepsilon}(\delta) = \text{clip}(\delta, -\varepsilon, \varepsilon).
\end{equation}

Following perturbation application, the combined waveform is normalized to prevent amplitude overflow. We employ a normalization scheme that preserves the relative contribution of the perturbation:
\begin{equation}
\tilde{x}_a^{\text{norm}} = \frac{\tilde{x}_a}{\max(|\tilde{x}_a|) + \epsilon},
\end{equation}
where $\epsilon = 10^{-8}$ is a small constant for numerical stability. This normalization ensures that the final audio remains within the valid amplitude range $[-1, 1]$ while maintaining the adversarial signal structure.

\subsection{Negative Language Modeling Loss}

The negative language modeling loss directly targets the model's output distribution by minimizing the probability assigned to the correct answer. Let $a^\star = (a_1^\star, a_2^\star, \ldots, a_m^\star)$ denote the ground-truth answer sequence of $m$ tokens. The standard cross-entropy language modeling loss under perturbed audio is:
\begin{equation}
\mathcal{L}_{\text{LM}}(\delta) = -\sum_{j=1}^{m} \log p_\theta(a_j^\star \mid a_{<j}^\star, \tilde{x}_a, x_v, q),
\end{equation}
where $p_\theta$ denotes the model's conditional probability distribution, $a_{<j}^\star$ represents the preceding tokens, $x_v$ is the video input, and $q$ is the question.

To suppress the correct answer, we negate this loss:
\begin{equation}
\begin{aligned}
\mathcal{L}_{\text{negLM}}(\delta)
&= -\mathcal{L}_{\text{LM}}(\delta) \\
&= \sum_{j=1}^{m} \log p_\theta\!\left(
a_j^\star \mid a_{<j}^\star, \tilde{x}_a, x_v, q
\right).
\end{aligned}
\end{equation}
Minimizing $\mathcal{L}_{\text{negLM}}$ is equivalent to minimizing the log-probability of the correct answer sequence. Intuitively, this objective pushes the model's probability mass away from the correct answer toward alternative responses.

This formulation connects to recent work on attacking large language models through projected gradient descent on input representations~\cite{geisler2025attackinglargelanguagemodels}. Unlike token-space attacks that require discrete optimization, our approach operates in the continuous audio waveform space where standard gradient methods apply directly.

\subsection{Encoder-Based Cosine Similarity Loss}

The encoder-based attack targets the audio representation space rather than the output distribution. Let $f_a: \mathbb{R}^{N_f \times 128} \rightarrow \mathbb{R}^{N_e \times d}$ denote the audio encoder that maps filterbank features to a sequence of $N_e$ embeddings of dimension $d$. We further apply a modality-specific projection layer $g_a: \mathbb{R}^{N_e \times d} \rightarrow \mathbb{R}^{N_e \times d_{\text{lm}}}$ that aligns audio embeddings with the language model's hidden dimension $d_{\text{lm}}$.

For clean audio $x_a$ with filterbank features $\mathbf{F}$ and perturbed audio $\tilde{x}_a$ with features $\tilde{\mathbf{F}}$, we compute the projected embeddings:
\begin{align}
\mathbf{E}_{\text{clean}} &= g_a(f_a(\mathbf{F})) \in \mathbb{R}^{N_e \times d_{\text{lm}}}, \\
\mathbf{E}_{\text{adv}} &= g_a(f_a(\tilde{\mathbf{F}})) \in \mathbb{R}^{N_e \times d_{\text{lm}}}.
\end{align}

The cosine similarity loss aggregates pairwise similarities across all embedding positions:
\begin{equation}
\mathcal{L}^{(\cos)}(\delta) = \frac{1}{N_e} \sum_{i=1}^{N_e} \frac{\mathbf{E}_{\text{clean}}[i] \cdot \mathbf{E}_{\text{adv}}[i]}{\|\mathbf{E}_{\text{clean}}[i]\|_2 \cdot \|\mathbf{E}_{\text{adv}}[i]\|_2},
\end{equation}
where $\mathbf{E}[i] \in \mathbb{R}^{d_{\text{lm}}}$ denotes the $i$-th embedding vector. Minimizing this loss pushes the adversarial embeddings to be orthogonal to their clean counterparts in the projected space.

The motivation for targeting the encoder rather than the output stems from the observation that encoder representations serve as the foundation for all downstream processing. By corrupting the audio representation at this early stage, we can induce failures that propagate through subsequent attention and decoding mechanisms without requiring supervision from specific output targets.

\subsection{Vision Attention Suppression Loss}

Attention mechanisms in transformer models determine how information flows between different input components. In multimodal models, cross-modal attention allows the model to ground its reasoning in relevant modalities. We design an objective that reduces the model's reliance on visual information by suppressing attention to vision tokens.

Let $\mathbf{A}^{(l,h)} \in \mathbb{R}^{N \times N}$ denote the attention matrix at layer $l \in \{1, \ldots, L\}$ and head $h \in \{1, \ldots, H\}$, where $N$ is the total sequence length including all modality tokens. Let $\mathcal{T}_v = \{v_{\text{start}}, v_{\text{start}}+1, \ldots, v_{\text{end}}\}$ denote the indices of vision tokens in the input sequence.

The total attention mass assigned to vision tokens is:
\begin{equation}
S_v(\delta) = \sum_{l=1}^{L} \sum_{h=1}^{H} \sum_{i=1}^{N} \sum_{j \in \mathcal{T}_v} \mathbf{A}^{(l,h)}_{i,j},
\end{equation}
where $\mathbf{A}^{(l,h)}_{i,j}$ represents the attention weight from position $i$ to position $j$. The vision attention suppression loss is simply:
\begin{equation}
\mathcal{L}^{(\text{visionatt})}(\delta) = S_v(\delta).
\end{equation}
Minimizing this loss reduces the total attention directed toward visual tokens across all layers and heads, effectively ``blinding'' the model to visual information even though the video input remains unchanged.

\subsection{Audio Attention Amplification Loss}

Complementary to suppressing visual attention, we design an objective that amplifies attention to audio tokens. The intuition is that by forcing the model to over-attend to the perturbed audio stream, we can maximize the influence of adversarial information on the model's reasoning process.

Let $\mathcal{T}_a = \{a_{\text{start}}, a_{\text{start}}+1, \ldots, a_{\text{end}}\}$ denote the indices of audio tokens. The total attention mass on audio tokens is:
\begin{equation}
S_a(\delta) = \sum_{l=1}^{L} \sum_{h=1}^{H} \sum_{i=1}^{N} \sum_{j \in \mathcal{T}_a} \mathbf{A}^{(l,h)}_{i,j}.
\end{equation}
To maximize this quantity, we define the loss as its negation:
\begin{equation}
\mathcal{L}^{(\text{audioatt})}(\delta) = -S_a(\delta).
\end{equation}
Minimizing $\mathcal{L}^{(\text{audioatt})}$ is equivalent to maximizing $S_a$, thereby increasing the model's reliance on audio tokens.

This objective is particularly effective when combined with encoder-space attacks. By simultaneously corrupting the audio representation and forcing the model to attend to it, we create a compounding effect where the model not only receives corrupted information but is also compelled to rely on it for reasoning.

\subsection{Attention Randomization Loss}

Beyond re-weighting attention across modalities, we consider an objective that directly disrupts the structure of attention patterns. The goal is to push attention matrices toward random configurations that lack the meaningful structure learned during training.

For each attention matrix $\mathbf{A}^{(l,h)}$, we construct a randomized target $\tilde{\mathbf{A}}^{(l,h)}$ as follows. Let $a_{\max} = \max(\mathbf{A}^{(l,h)})$ and $a_{\min} = \min(\mathbf{A}^{(l,h)})$ denote the range of attention logits. We sample a random matrix:
\begin{equation}
\mathbf{R}^{(l,h)} = (a_{\max} - a_{\min}) \cdot \mathbf{U} + a_{\min},
\end{equation}
where $\mathbf{U} \in \mathbb{R}^{N \times N}$ has entries sampled uniformly from $[0, 1]$. To preserve the autoregressive structure of causal attention, we apply a lower-triangular mask:
\begin{equation}
\tilde{\mathbf{A}}^{(l,h)} = \mathbf{R}^{(l,h)} \odot \mathbf{M}_{\text{tril}},
\end{equation}
where $\mathbf{M}_{\text{tril}}$ is a lower-triangular matrix of ones and $\odot$ denotes element-wise multiplication.

The attention randomization loss measures the divergence between actual and randomized attention using KL divergence over the softmax-normalized distributions:
\begin{equation}
\begin{aligned}
\mathcal{L}^{(\text{randatt})}(\delta)
&= \sum_{l=1}^{L} \sum_{h=1}^{H}
D_{\text{KL}}\!\Big(
\text{softmax}(\mathbf{A}^{(l,h)})
\,\Big\|\, \\
&\qquad \text{softmax}(\tilde{\mathbf{A}}^{(l,h)})
\Big),
\end{aligned}
\end{equation}
where $D_{\text{KL}}(P \| Q) = \sum_i P_i \log(P_i / Q_i)$ is the Kullback-Leibler divergence and the softmax is applied row-wise.

Minimizing this loss pushes the attention patterns toward the randomized targets. Since the targets are sampled independently at each forward pass, this objective effectively encourages attention patterns that are uniformly random rather than structured around semantically meaningful relationships.

\subsection{Hidden-State Similarity Loss}

The hidden-state attack targets the internal representations at each transformer layer. Unlike the encoder-based attack which operates only on the audio encoder outputs, this objective propagates through the entire model architecture.

Let $\mathbf{h}_l(x) \in \mathbb{R}^{N \times d_{\text{lm}}}$ denote the hidden states at layer $l$ for input $x$. For clean input $(x_a, x_v, q)$ and perturbed input $(\tilde{x}_a, x_v, q)$, we compute hidden states at each layer and measure their similarity using cosine distance:
\begin{equation}
\begin{aligned}
\mathcal{L}^{(\text{hidden-cos})}(\delta)
&= \frac{1}{L} \sum_{l=1}^{L} \frac{1}{N} \sum_{i=1}^{N} \\
&\qquad \cos\Big(\mathbf{h}_l^{(i)}(\tilde{x}_a),\, \mathbf{h}_l^{(i)}(x_a)\Big),
\end{aligned}
\end{equation}
where $\mathbf{h}_l^{(i)} \in \mathbb{R}^{d_{\text{lm}}}$ denotes the hidden state at position $i$ in layer $l$.

This formulation requires computing a full forward pass for both clean and perturbed inputs, then comparing their hidden representations position-by-position and layer-by-layer. Minimizing this loss pushes the adversarial hidden states to diverge from their clean counterparts throughout the network depth.

The hidden-state attack is motivated by work on representation engineering and activation steering in language models~\cite{shu2025layerwiseperturbationssparseautoencoders}, which demonstrates that manipulating internal representations can induce specific behavioral changes. By targeting hidden states, we can potentially induce more fundamental failures in the model's reasoning process compared to output-level attacks.

\subsection{Combined Loss}

To leverage the complementary failure modes induced by different attack objectives, we combine all losses into a unified objective:
\begin{equation}
\begin{aligned}
\mathcal{L}^{(\text{combined})}(\delta)
&= \mathcal{L}_{\text{negLM}}(\delta) + \mathcal{L}^{(\cos)}(\delta) \\
&\quad + \mathcal{L}^{(\text{visionatt})}(\delta) + \mathcal{L}^{(\text{audioatt})}(\delta) \\
&\quad + \mathcal{L}^{(\text{randatt})}(\delta) + \mathcal{L}^{(\text{hidden-cos})}(\delta).
\end{aligned}
\end{equation}
All loss components are combined with equal weighting. We do not tune individual loss weights, as our goal is to assess whether a simple combination of diverse objectives can outperform individual attacks.

The combined loss aggregates signals from multiple stages of the audio-to-output pipeline: encoder representations, cross-modal attention allocation, internal hidden states, and output likelihood. This multi-view formulation provides gradient information that can guide the perturbation toward configurations that simultaneously disrupt multiple computational mechanisms.

\section{Optimization Algorithm}
\label{appendix:algorithm}

This section presents the complete optimization procedure for learning adversarial audio perturbations. We decompose the attack into three modular algorithms: audio preprocessing (Algorithm~\ref{alg:preprocess}), single-step gradient update (Algorithm~\ref{alg:gradient_step}), and the main optimization loop (Algorithm~\ref{alg:main_loop}).

\subsection{Audio Preprocessing}

Algorithm~\ref{alg:preprocess} describes the audio preprocessing pipeline that transforms raw waveforms into model-ready features. This procedure handles perturbation application, amplitude normalization, and mel-filterbank extraction.

\begin{algorithm}[ht]
\caption{Audio Preprocessing Pipeline}
\label{alg:preprocess}
\begin{algorithmic}[1]
\REQUIRE Clean audio $x_a \in \mathbb{R}^T$, perturbation $\delta \in \mathbb{R}^{T_\delta}$
\ENSURE Mel-filterbank features $\mathbf{F}$
\STATE \textbf{// Cyclic perturbation extension}
\FOR{$t = 0$ to $T-1$}
    \STATE $\tilde{x}_a[t] \gets x_a[t] + \delta[t \mod T_\delta]$
\ENDFOR
\STATE \textbf{// Amplitude normalization}
\STATE $\tilde{x}_a^{\text{norm}} \gets \tilde{x}_a \,/\, (\max_{t}|\tilde{x}_a[t]| + 10^{-8})$
\STATE \textbf{// Scale to 16-bit range}
\STATE $w \gets \tilde{x}_a^{\text{norm}} \cdot 2^{15}$
\STATE \textbf{// Mel-filterbank extraction}
\STATE $\mathbf{F} \gets \textsc{MelFilterbank}(w, \text{bins}=128, f_s=16\text{kHz})$
\RETURN $\mathbf{F}$, $\tilde{x}_a^{\text{norm}}$
\end{algorithmic}
\end{algorithm}

The mel-filterbank computation uses the Kaldi library with 128 mel bins, 25ms frame length (400 samples), and 10ms frame shift (160 samples) at 16kHz sampling rate. The entire pipeline is differentiable, enabling gradient flow from the loss function to the raw perturbation.

\subsection{Gradient Update Step}

Algorithm~\ref{alg:gradient_step} describes a single gradient update step for the perturbation. This modular formulation separates the gradient computation and projection from the outer optimization loop.

\begin{algorithm}[ht]
\caption{Single-Step Gradient Update}
\label{alg:gradient_step}
\begin{algorithmic}[1]
\REQUIRE Current perturbation $\delta$, sample $(x_a, x_v, q, a^\star)$
\REQUIRE Model $M_\theta$, loss function $\mathcal{L}$, budget $\varepsilon$
\REQUIRE Gradient scaling factor $\gamma$, optimizer state
\ENSURE Updated perturbation $\delta$, sample loss $\ell$
\STATE $\mathbf{F}, \tilde{x}_a \gets \textsc{Preprocess}(x_a, \delta)$ \COMMENT{Algorithm~\ref{alg:preprocess}}
\STATE \textbf{// Forward pass through frozen model}
\STATE $\text{outputs} \gets M_\theta(\mathbf{F}, x_v, q)$
\STATE \textbf{// Compute adversarial loss}
\STATE $\ell \gets \mathcal{L}(\text{outputs}, a^\star, x_a, \tilde{x}_a)$
\STATE \textbf{// Backward pass}
\STATE $\mathbf{g} \gets \nabla_\delta \ell$
\STATE \textbf{// Handle numerical instabilities}
\STATE $\mathbf{g}[\text{isnan}(\mathbf{g})] \gets 0$
\STATE \textbf{// Apply gradient scaling}
\STATE $\mathbf{g} \gets \mathbf{g} \cdot \gamma$
\STATE \textbf{// Adam update}
\STATE $\delta \gets \textsc{AdamUpdate}(\delta, \mathbf{g}, \text{optimizer state})$
\STATE \textbf{// Project onto constraint set}
\STATE $\delta \gets \text{clip}(\delta, -\varepsilon, \varepsilon)$
\RETURN $\delta$, $\ell$
\end{algorithmic}
\end{algorithm}

The gradient scaling factor $\gamma$ stabilizes optimization across different loss functions. We use $\gamma \in \{1, 10^{-5}, 10^{4}\}$ depending on the magnitude of raw gradients. 

\subsection{Main Optimization Loop}

Algorithm~\ref{alg:main_loop} describes the outer optimization loop with early stopping. The algorithm iterates over the training dataset, accumulating gradients and tracking the best perturbation found during training.

\begin{algorithm}[ht]
\caption{Main Attack Optimization Loop}
\label{alg:main_loop}
\begin{algorithmic}[1]
\REQUIRE Dataset $\mathcal{D} = \{(x_a^{(i)}, x_v^{(i)}, q^{(i)}, a^{\star(i)})\}_{i=1}^{N}$
\REQUIRE Model $M_\theta$, loss $\mathcal{L}$, budget $\varepsilon$, learning rate $\eta$
\REQUIRE Max epochs $K$, patience $P$
\ENSURE Optimized perturbation $\delta^{\text{best}}$
\STATE \textbf{// Initialize perturbation from natural audio}
\STATE $\delta \gets \text{clip}(\text{BellSound}, -\varepsilon, \varepsilon)$
\STATE \textbf{// Initialize Adam optimizer}
\STATE optimizer $\gets \textsc{Adam}(\eta, \beta_1=0.9, \beta_2=0.999)$
\STATE best\_loss $\gets \infty$
\STATE epochs\_no\_improve $\gets 0$
\FOR{epoch $= 1$ to $K$}
    \STATE epoch\_loss $\gets 0$
    \STATE \textbf{// Iterate over training samples}
    \FOR{$(x_a, x_v, q, a^\star) \in \mathcal{D}$}
        \STATE $\delta, \ell \gets \textsc{GradientStep}(\delta, (x_a, x_v, q, a^\star))$
        \STATE epoch\_loss $\gets$ epoch\_loss $+ \ell$
    \ENDFOR
    \STATE epoch\_loss $\gets$ epoch\_loss $/\, |\mathcal{D}|$
    \STATE \textbf{// Track best perturbation}
    \IF{epoch\_loss $<$ best\_loss}
        \STATE best\_loss $\gets$ epoch\_loss
        \STATE $\delta^{\text{best}} \gets \delta$
        \STATE epochs\_no\_improve $\gets 0$
    \ELSE
        \STATE epochs\_no\_improve $\gets$ epochs\_no\_improve $+ 1$
    \ENDIF
    \STATE \textbf{// Early stopping check}
    \IF{epochs\_no\_improve $\geq P$}
        \STATE \textbf{break}
    \ENDIF
\ENDFOR
\RETURN $\delta^{\text{best}}$
\end{algorithmic}
\end{algorithm}

Training terminates when the loss does not improve for $P$ consecutive epochs. We use $P = 10$ in most experiments. This criterion prevents overfitting to the training set and reduces computational cost for attacks that converge quickly. The best perturbation is saved to disk after each improvement for recovery in case of interruption.

\clearpage
\section{Implementation Details}
\label{appendix:implementation}

\subsection{Optimization Hyperparameters}

All attacks use projected gradient descent with the Adam optimizer. Table~\ref{tab:hyperparameters} summarizes the hyperparameter settings used across experiments.

\begin{table}[ht]
\centering
\adjustbox{max width=\linewidth}{
\begin{tabular}{lc}
\toprule
\textbf{Hyperparameter} & \textbf{Value} \\
\midrule
Learning rate $\eta$ & $10^{-4}$ \\
Adam $\beta_1$ & 0.9 \\
Adam $\beta_2$ & 0.999 \\
Adam $\epsilon$ & $10^{-8}$ \\
Batch size & 1 \\
Maximum epochs & 150 \\
Early stopping patience & 10 epochs \\
Gradient clipping norm & 1.0 \\
Attack budget $\varepsilon$ & $\{0.3, 0.5, 0.7, 1.0\}$ \\
Perturbation length & 10 seconds \\
Sampling rate & 16 kHz \\
\bottomrule
\end{tabular}
}
\caption{Hyperparameter settings for adversarial attack optimization.}
\label{tab:hyperparameters}
\end{table}

\subsection{Perturbation Initialization}

The choice of initialization for the perturbation $\delta^{(0)}$ affects both convergence speed and final attack quality. We experimented with two initialization strategies:

\textit{(1) Random initialization:} Each element of $\delta^{(0)}$ is sampled uniformly from $[-\varepsilon, \varepsilon]$. This provides a generic starting point but may require more iterations to converge.

\textit{(2) Natural audio initialization:} We initialize $\delta^{(0)}$ using the audio track from a natural video in the dataset. Specifically, we use a recording of bells ringing from the AVQA dataset, clipped to the attack budget $\varepsilon$. This sample is removed from the training set to prevent data leakage. We found that natural audio initialization leads to perturbations that sound more natural when played independently. 

Unless specified otherwise, we use natural audio initialization.

\subsection{Convergence Conditions}
\label{app:convergence}

For the attack budget analysis experiments, we allow the optimization to run until convergence rather than fixing a maximum number of epochs. Convergence is defined as no improvement in the training loss for 10 consecutive epochs. Under this criterion, different attack budgets require different numbers of epochs to converge, providing insight into the optimization landscape at different perturbation magnitudes.

\subsection{Learning Rate Scheduling}

We implement two learning rate scheduling strategies depending on the experiment:

\textit{(1) Reduce-on-plateau:} The learning rate is reduced by a factor of 0.1 when the loss does not improve for a specified patience window. This is used for experiments where we train until convergence.

\textit{(2) Cosine annealing with warmup:} For fixed-epoch experiments, we use a warmup phase followed by cosine annealing. The learning rate increases linearly during warmup, then follows a cosine decay schedule:
\begin{equation}
\eta_t = \begin{cases}
\eta \cdot \frac{t}{t_{\text{warm}}} & \text{if } t < t_{\text{warm}} \\
\frac{\eta}{2}\left(1 + \cos\left(\pi \cdot \frac{t - t_{\text{warm}}}{T - t_{\text{warm}}}\right)\right) & \text{otherwise}
\end{cases}
\end{equation}
where $t_{\text{warm}}$ is the warmup duration and $T$ is the total number of training steps.

Unless specified otherwise, we use reduce-on-plateau.

\subsection{Hardware and Computational Resources}

All experiments were conducted on NVIDIA GPUs including H200, A100, L40s, and A40s models. The choice of GPU depended on availability and memory requirements of specific experiments. VideoLLAMA2 experiments used full precision (FP32) to ensure numerical stability during gradient computation through the audio preprocessing pipeline. Qwen model experiments used mixed precision (FP16) to reduce memory consumption and accelerate training.

The total computational budget for all experiments reported in this paper is approximately 2,500 GPU-hours. Individual attack training runs range from 20-25 GPU-hours depending on the dataset size, number of epochs, and model complexity. Evaluation on held-out test sets requires additional compute for inference but is substantially faster than training.

\subsection{Reproducibility}

All experiments use fixed random seeds for reproducibility. We set:

\textit{(1)} NumPy random seed: 42

\textit{(2)} PyTorch CUDA seed: 30

These settings ensure that perturbation initialization, data shuffling, and GPU operations produce consistent results across runs. We found that attack success rates are sensitive to random initialization, making seed fixing essential for reproducible comparisons.

\section{Evaluation Metrics}
\label{appendix:eval_metrics}

This section provides detailed definitions and interpretations for all evaluation metrics used in our experiments.

\subsection{Attack Success Rate}

The Attack Success Rate (ASR) quantifies the fraction of originally correct predictions that are flipped to incorrect predictions after applying the adversarial perturbation. This metric specifically measures the attack's ability to induce failures on samples where the model would otherwise succeed, excluding samples where the model was already incorrect on clean inputs.

Formally, given an evaluation dataset $\mathcal{D} = \{(x_a^{(i)}, x_v^{(i)}, q^{(i)}, a^{\star(i)})\}_{i=1}^{N}$, we define:
\begin{equation}
\text{CleanCorrect}^{(i)} = \mathbf{1}\left[\hat{a}(x_a^{(i)}, x_v^{(i)}, q^{(i)}; \theta) = a^{\star(i)}\right],
\end{equation}
where $\hat{a}(\cdot)$ denotes the model's predicted answer and $\mathbf{1}[\cdot]$ is the indicator function. Similarly, for perturbed audio $\tilde{x}_a^{(i)} = x_a^{(i)} + \delta$:
\begin{equation}
\begin{aligned}
\text{AttackSuccess}^{(i)}
&= \mathbf{1}\Big[
\text{CleanCorrect}^{(i)} = 1 \;\wedge \\
&\qquad \hat{a}(\tilde{x}_a^{(i)}, x_v^{(i)}, q^{(i)}; \theta) \neq a^{\star(i)}
\Big].
\end{aligned}
\end{equation}
The ASR is then computed as:
\begin{equation}
\text{ASR} = \frac{\sum_{i=1}^{N} \text{AttackSuccess}^{(i)}}{\sum_{i=1}^{N} \text{CleanCorrect}^{(i)}}.
\end{equation}

This definition ensures that ASR measures genuine attack-induced failures rather than being inflated by baseline model errors. An ASR of 100\% indicates that the attack successfully flips all correct predictions to incorrect ones.

\subsection{LPIPS for Audio}

We adapt the Learned Perceptual Image Patch Similarity (LPIPS) metric to the audio domain by computing it on log-mel spectrogram representations. Given clean audio $x_a$ and perturbed audio $\tilde{x}_a$, we first compute their log-mel spectrograms $\mathbf{S}$ and $\tilde{\mathbf{S}}$ respectively. These spectrograms are treated as single-channel images and fed to a pretrained neural network (AlexNet or VGG) to extract deep features. 

LPIPS computes the distance between activations at multiple layers of the network:
\begin{equation}
\begin{aligned}
\text{LPIPS}(x_a, \tilde{x}_a)
&= \sum_{l} \frac{1}{H_l W_l} \sum_{h,w} \left\| \Delta_{l,hw} \right\|_2^2, \\
\Delta_{l,hw}
&= w_l \odot \Big(\phi_l(\mathbf{S})_{hw} - \phi_l(\tilde{\mathbf{S}})_{hw}\Big),
\end{aligned}
\end{equation}
where $\phi_l$ extracts features at layer $l$, $w_l$ are learned weights, and the sum is over spatial positions $(h, w)$ in the feature maps of height $H_l$ and width $W_l$.

Unlike pixel-wise metrics such as MSE, LPIPS captures perceptual differences by comparing high-level features. Lower LPIPS values indicate higher perceptual similarity between clean and perturbed audio, while higher values correspond to increasingly noticeable distortions. This comparison is motivated by ~\cite{chary2025spectrogrampatchcodec2d} where authors use LPIPS for audio reconstruction.

\subsection{Scale-Invariant Signal-to-Noise Ratio}

Scale-Invariant Signal-to-Noise Ratio (SI-SNR) measures the similarity between a reference signal and an estimated signal while removing any global gain differences. This property makes SI-SNR more robust than standard SNR for evaluating audio perturbations, as it focuses on waveform shape rather than overall amplitude.

Given clean signal $\mathbf{x} \in \mathbb{R}^T$ and perturbed signal $\hat{\mathbf{x}} \in \mathbb{R}^T$, SI-SNR is computed as follows. First, both signals are zero-meaned:
\begin{equation}
\mathbf{x} \gets \mathbf{x} - \frac{1}{T}\sum_{t=1}^{T} x_t, \quad \hat{\mathbf{x}} \gets \hat{\mathbf{x}} - \frac{1}{T}\sum_{t=1}^{T} \hat{x}_t.
\end{equation}
Next, we project $\hat{\mathbf{x}}$ onto $\mathbf{x}$ to obtain the signal-aligned component:
\begin{equation}
\mathbf{x}_{\text{target}} = \frac{\langle \hat{\mathbf{x}}, \mathbf{x} \rangle}{\|\mathbf{x}\|^2} \mathbf{x}.
\end{equation}
The residual error orthogonal to $\mathbf{x}$ is:
\begin{equation}
\mathbf{e}_{\text{noise}} = \hat{\mathbf{x}} - \mathbf{x}_{\text{target}}.
\end{equation}
Finally, SI-SNR is the logarithmic ratio of signal energy to noise energy:
\begin{equation}
\text{SI-SNR}(\hat{\mathbf{x}}, \mathbf{x}) = 10 \log_{10} \left( \frac{\|\mathbf{x}_{\text{target}}\|^2}{\|\mathbf{e}_{\text{noise}}\|^2} \right).
\end{equation}

Higher SI-SNR values indicate closer alignment between perturbed and clean signals. A value of 0 dB means the noise energy equals the signal energy. Negative values indicate that the perturbation dominates the original signal, though the original audio may still be audible if the perturbation is structured rather than random.

\subsection{Word Error Rate}

For evaluating attacks on automatic speech recognition systems, we use Word Error Rate (WER), the standard metric for transcription accuracy:
\begin{equation}
\text{WER} = \frac{S + D + I}{N} \times 100\%,
\end{equation}
where $S$ is the number of word substitutions, $D$ is the number of deletions, $I$ is the number of insertions required to transform the predicted transcript into the reference, and $N$ is the total number of words in the reference transcript.

We report both absolute WER on attacked audio and the change in WER ($\Delta$WER = WER(attacked) $-$ WER(clean)) to quantify attack impact relative to baseline performance.

\section{LLM-as-a-Judge Evaluation}
\label{appendix:llm_judge}

For evaluating model predictions on the AVSD dataset, we employ an LLM-as-a-judge approach that automates the assessment of free-form text generation quality. This section describes the evaluation protocol and provides the complete prompt template.

\subsection{Evaluation Protocol}

The evaluation proceeds through the following steps:

\textit{(1)} For each sample in the AVSD validation set, we collect the ground truth caption, summary, and all conversation Q\&A turns that describe the video content.

\textit{(2)} We generate predictions using both clean audio and adversarially perturbed audio, obtaining two descriptions for each video.

\textit{(3)} Each prediction is independently evaluated by the LLM judge against the ground truth information. The judge assesses factual accuracy, consistency with visual content, and absence of hallucinations.

\textit{(4)} The judge outputs a structured response containing: a brief analysis explaining the reasoning, a binary verdict (CORRECT or INCORRECT), and a confidence level (HIGH, MEDIUM, or LOW).

\textit{(5)} We compute the attack success rate as the fraction of samples where the clean prediction was judged CORRECT but the adversarial prediction was judged INCORRECT.

This automated evaluation enables consistent and scalable assessment across the full validation set while maintaining alignment with human judgment criteria. The structured output format allows for reliable parsing of verdicts without manual inspection.

\subsection{Prompt Template}

Figure~\ref{fig:llm_judge_prompt} presents the complete prompt template used for LLM-as-a-judge evaluation.

\begin{figure*}[ht]
\begin{tcolorbox}[colback=teallight, colframe=tealdark, title=\textbf{LLM Judge Prompt Template}, width=\textwidth, breakable]
\small
\texttt{You are an expert evaluator for video description quality assessment. Your task is to determine whether a model's prediction accurately describes a video based on ground truth information.}

\vspace{0.3cm}
\textbf{=== GROUND TRUTH INFORMATION ===}

\texttt{**Video Caption:** \{gt\_caption\}}

\texttt{**Video Summary:** \{gt\_summary\}}

\texttt{**Detailed Q\&A about the video:**}\\
\texttt{\{conversation\_context\}}

\vspace{0.3cm}
\textbf{=== MODEL PREDICTION TO EVALUATE ===}\\
\texttt{\{prediction\}}

\vspace{0.3cm}
\textbf{=== EVALUATION CRITERIA ===}

\texttt{Evaluate the prediction on these dimensions:}
\begin{enumerate}
\item \textbf{Subject Identification}: Does the prediction correctly identify the main person/people and their gender/age if mentioned?
\item \textbf{Action Accuracy}: Does the prediction correctly describe what the person is doing (actions, activities)?
\item \textbf{Object Recognition}: Are the objects mentioned in the prediction consistent with ground truth (e.g., food items, furniture, clothing)?
\item \textbf{Setting/Location}: Is the location/setting correctly identified (room type, indoor/outdoor)?
\item \textbf{Factual Consistency}: Are there any direct contradictions or hallucinations compared to ground truth?
\end{enumerate}

\vspace{0.3cm}
\textbf{=== JUDGMENT GUIDELINES ===}

\texttt{Mark as \textbf{CORRECT} if:}
\begin{itemize}
\item The prediction captures the essential scene accurately
\item Main actions and subjects are correctly identified
\item No major factual errors or hallucinations
\item Minor omissions or paraphrasing are acceptable
\end{itemize}

\texttt{Mark as \textbf{INCORRECT} if:}
\begin{itemize}
\item The prediction misidentifies the main subject or their actions
\item There are factual contradictions with ground truth
\item The prediction describes a substantially different scene
\item Critical details are wrong (e.g., wrong objects, wrong location type)
\end{itemize}

\vspace{0.3cm}
\textbf{=== OUTPUT FORMAT ===}

\texttt{Provide your response in this EXACT format:}

\texttt{ANALYSIS: [Briefly explain your reasoning in 2-3 sentences, comparing key elements]}

\texttt{VERDICT: [CORRECT/INCORRECT]}

\texttt{CONFIDENCE: [HIGH/MEDIUM/LOW]}
\end{tcolorbox}
\caption{Prompt template used for LLM-as-a-judge evaluation on AVSD. The judge evaluates predictions against ground truth captions, summaries, and conversation context.}
\label{fig:llm_judge_prompt}
\end{figure*}

\subsection{Judge Model Selection}

We use the \texttt{openai/gpt-oss-20b} model as the evaluation judge. This model was selected for its strong instruction-following capabilities and consistent output formatting. The temperature is set to 0 to ensure deterministic evaluations across samples.

\clearpage
\section{Additional Results and Numerical Tables}
\label{appendix:additional_results}

This section provides exact numerical values for all figures and additional analysis presented in the main paper.

\subsection{Training Iterations and Attack Success Rate}

Table~\ref{tab:iterations_vs_asr} reports the relationship between total training iterations, dataset size, and attack success rate for $\mathcal{L}_{\text{negLM}}$. These results correspond to Figure~\ref{plot:iteration_vs_asr_loss1} in the main paper.

\begin{table}[ht]
\centering
\adjustbox{max width=\linewidth}{
\begin{tabular}{ccc}
\toprule
\textbf{Total Iterations} & \textbf{ASR (\%)} & \textbf{Dataset Size}\\
\midrule
200k & 1.93 & 20k \\
400k & 73.64 & 20k\\
100k & 2.16 & 10k\\
200k & 78.82 & 10k\\
500k & 93.15 & 10k\\
300k & 10.27 & 2k \\
121k & 36.76 & 1k \\
200k & 64.07 & 1k \\
372k & 87.00 & 1k \\
\bottomrule
\end{tabular}
}
\caption{Training iterations, dataset size, and attack success rate for $\mathcal{L}_{\text{negLM}}$ on VideoLLAMA2. Extended optimization on smaller datasets yields higher ASR than brief training on larger datasets.}
\label{tab:iterations_vs_asr}
\end{table}

The results reveal that attack effectiveness depends more on total optimization iterations than on dataset diversity. Configurations with only 1,000 training samples but 372,000 iterations achieve 87\% ASR, substantially outperforming configurations with 20,000 samples but only 200,000 iterations (1.93\% ASR). This pattern suggests that adversarial perturbations exploit consistent model behaviors that can be identified through repeated optimization on a small, representative subset of the data.

\subsection{Perceptibility Analysis}

Table~\ref{tab:perceptibility_full} reports perceptual distortion metrics alongside attack effectiveness for each objective. These values correspond to Figure~\ref{plot:asr_vs_impercetibility}.

\begin{table}[ht]
\centering
\adjustbox{max width=\linewidth}{
\begin{tabular}{lccc}
\toprule
\textbf{Attack Objective} & \textbf{$\Delta$WER} & \textbf{LPIPS} & \textbf{SI-SNR (dB)}\\
\midrule
$\mathcal{L}_\text{negLM}$ & 0.972 & 0.22 & $-11.48$\\
$\mathcal{L}^{(\text{cos})}$ & 0.959 & 0.08 & $-1.77$ \\
$\mathcal{L}^{(\text{visionatt})}$ & 0.100 & 0.07 & $-1.02$ \\
$\mathcal{L}^{(\text{audioatt})}$ & 0.047 & 0.06 & 0.33 \\
$\mathcal{L}^{(\text{randatt})}$ & 0.062 & 0.06 & 0.05\\
$\mathcal{L}^{(\text{hidden-cos})}$ & 0.036 & 0.06 & 0.56\\
$\mathcal{L}^{(\text{combined})}$ & 0.491 & 0.14 & $-6.23$\\
\bottomrule
\end{tabular}
}
\caption{Perceptual distortion metrics and WER difference for each attack objective. Higher $\Delta$WER correlates with higher LPIPS, indicating that speech recognition models respond primarily to acoustic distortion magnitude.}
\label{tab:perceptibility_full}
\end{table}

The encoder-space attack $\mathcal{L}^{(\cos)}$ achieves a favorable trade-off between attack effectiveness and perceptibility. Despite inducing nearly as much damage to Whisper as $\mathcal{L}_{\text{negLM}}$ (0.959 vs 0.972 $\Delta$WER), it maintains substantially lower perceptual distortion (LPIPS 0.08 vs 0.22, SI-SNR $-1.77$ vs $-11.48$ dB). This indicates that the encoder-space attack exploits structured vulnerabilities in the audio representation rather than relying on brute-force signal corruption.

\subsection{Attention Allocation Analysis}

Table~\ref{tab:attn_per_token} reports the average attention mass assigned to audio and video tokens under different attack conditions. Values are normalized by the number of layers and tokens to enable comparison across models.

\begin{table}[ht]
\centering
\adjustbox{max width=\linewidth}{
\begin{tabular}{lcccc}
\toprule
\textbf{Objective} & \textbf{VL2-audio} & \textbf{VL2-video} & \textbf{Q2.5-audio} & \textbf{Q2.5-video} \\
\midrule
Clean & 9.01 & \textbf{15.50} & 13.01 & \textbf{39.04} \\
$\mathcal{L}_\text{negLM}$ & 9.55 & 14.94 & 12.96 & 31.86\\
$\mathcal{L}^{(\text{cos})}$ & 13.55 & 13.96 & 13.03 & 35.00\\
$\mathcal{L}^{(\text{visionatt})}$ & 14.03 & 13.49 & \textbf{13.50} & 34.42\\
$\mathcal{L}^{(\text{audioatt})}$ & 21.49 & 13.85 & 12.88 & 32.73\\
$\mathcal{L}^{(\text{randatt})}$ & 10.51 & 14.65 & 13.03 & 32.48\\
$\mathcal{L}^{(\text{hidden-cos})}$ & 13.55 & 14.43 & 13.04 & 34.51\\
$\mathcal{L}^{(\text{combined})}$ & \textbf{29.23} & 13.30 & 13.17 & 34.34\\
\bottomrule
\end{tabular}
}
\caption{Average attention logits assigned to audio and video tokens for VideoLLAMA2 (VL2) and Qwen 2.5 Omni (Q2.5), normalized by layer and token count. The combined attack produces the largest shift in audio-token attention on VideoLLAMA2 (29.23 vs 9.01 clean).}
\label{tab:attn_per_token}
\end{table}

\subsection{Sequence Confidence Under Attack}

Table~\ref{tab:seq_confidence_full} reports the average sequence confidence of generated outputs under clean and adversarial conditions. Confidence is computed as the exponential of the mean log-probability of all generated tokens.

\begin{table*}[ht]
\centering
\adjustbox{max width=\textwidth}{
\begin{tabular}{lccccccccc}
\toprule
\textbf{Objective} & \textbf{AVQA-Adv} & \textbf{AVQA-Clean} & \textbf{$\Delta$} & \textbf{M-AVQA-Clean} & \textbf{M-AVQA-Adv} & \textbf{$\Delta$} & \textbf{AVSD-Adv} & \textbf{AVSD-Clean} & \textbf{$\Delta$} \\
\midrule
$\mathcal{L}_\text{negLM}$ & 0.756 & 0.816 & 0.060 & 0.752 & 0.818 & 0.066 & 0.640 & 0.650 & 0.010 \\
$\mathcal{L}^{(\text{cos})}$ & 0.800 & 0.930 & 0.130 & 0.741 & 0.823 & 0.082 & 0.650 & 0.660 & 0.010 \\
$\mathcal{L}^{(\text{visionatt})}$ & 0.838 & 0.935 & 0.097 & 0.765 & 0.798 & 0.033 & 0.660 & 0.650 & $-0.010$ \\
$\mathcal{L}^{(\text{audioatt})}$ & 0.712 & 0.935 & 0.223 & 0.770 & 0.813 & 0.043 & 0.660 & 0.650 & $-0.010$ \\
$\mathcal{L}^{(\text{randatt})}$ & 0.713 & 0.798 & 0.085 & 0.770 & 0.816 & 0.046 & 0.650 & 0.650 & 0.000 \\
$\mathcal{L}^{(\text{hidden-cos})}$ & 0.822 & 0.930 & 0.108 & 0.737 & 0.800 & 0.063 & 0.650 & 0.640 & $-0.010$ \\
$\mathcal{L}^{(\text{combined})}$ & 0.850 & 0.930 & 0.080 & 0.750 & 0.855 & 0.105 & 0.642 & 0.650 & 0.008 \\
\bottomrule
\end{tabular}
}
\caption{Sequence confidence for clean and adversarial settings across datasets, computed for successful attacks only. $\Delta$ = Clean $-$ Adv. Small or negative differences indicate that adversarial failures arise from internal misalignment rather than model uncertainty.}
\label{tab:seq_confidence_full}
\end{table*}

The small confidence differences under attack are notable. For example, the combined attack on AVQA achieves 96.03\% ASR while the confidence difference is only 0.080 (0.930 clean vs 0.850 adversarial). This indicates that the model produces incorrect answers with nearly the same confidence as correct answers, making adversarial failures difficult to detect through output uncertainty monitoring.

\subsection{Clean versus Attacked Accuracy}

Table~\ref{tab:accuracy} reports model accuracy on clean and adversarial audio across AVQA and Music-AVQA for all attack objectives.

\begin{table*}[t]
\centering
\adjustbox{max width=\textwidth}{
\begin{tabular}{lcccccc}
\toprule
\textbf{Objective} & \textbf{AVQA Clean} & \textbf{AVQA Attack} & \textbf{$\Delta$} & \textbf{Music-AVQA Clean} & \textbf{Music-AVQA Attack} & \textbf{$\Delta$}\\
\midrule
$\mathcal{L}_\text{negLM}$ & 0.956 & 0.850 & 0.106 & 0.807 & 0.743 & 0.064\\
$\mathcal{L}^{(\text{cos})}$ & 0.956 & 0.110 & 0.846 & 0.807 & 0.747 & 0.060\\
$\mathcal{L}^{(\text{visionatt})}$ & 0.956 & 0.920 & 0.036 & 0.807 & 0.777 & 0.030\\
$\mathcal{L}^{(\text{audioatt})}$ & 0.956 & 0.410 & 0.546 & 0.807 & 0.769 & 0.038\\
$\mathcal{L}^{(\text{randatt})}$ & 0.956 & 0.910 & 0.046 & 0.807 & 0.758 & 0.049\\
$\mathcal{L}^{(\text{hidden-cos})}$ & 0.956 & 0.795 & 0.161 & 0.807 & 0.766 & 0.041\\
$\mathcal{L}^{(\text{combined})}$ & 0.956 & 0.039 & \textbf{0.918} & 0.807 & 0.750 & \textbf{0.057}\\
\bottomrule
\end{tabular}
}
\caption{Accuracy comparison between clean and adversarial audio. $\Delta$ denotes the accuracy drop (Clean $-$ Attack). High ASR is driven by substantial attack-induced degradation, not low baseline accuracy.}
\label{tab:accuracy}
\end{table*}

The results confirm that high ASR values reflect genuine attack-induced failures rather than poor baseline performance. The model achieves 95.6\% accuracy on clean AVQA inputs. Under the combined attack, accuracy drops to 3.9\%, representing a 91.8 percentage point degradation that is attributable entirely to the adversarial perturbation.

\subsection{Black-Box Transfer to Whisper}

Table~\ref{tab:whisper_results} reports the transferability of audio-only adversarial perturbations to Whisper Large-v2 in a black-box setting, evaluated on LibriSpeech.

\begin{table}[ht]
\centering
\adjustbox{max width=\linewidth}{
\begin{tabular}{lcc}
\toprule
\textbf{Attack Objective} & \textbf{ASR\% on Whisper} & \textbf{$\Delta$WER}\\
\midrule
$\mathcal{L}_\text{negLM}$ & \textbf{98.20} & 0.972 \\
$\mathcal{L}^{(\text{cos})}$ & 97.38 & 0.959 \\
$\mathcal{L}^{(\text{visionatt})}$ & 17.99 & 0.100 \\
$\mathcal{L}^{(\text{audioatt})}$ & 9.25 & 0.047 \\
$\mathcal{L}^{(\text{randatt})}$ & 11.29 & 0.062\\
$\mathcal{L}^{(\text{hidden-cos})}$ & 7.37 & 0.036\\
$\mathcal{L}^{(\text{combined})}$ & 66.88 & 0.491\\
\bottomrule
\end{tabular}
}
\caption{Black-box transfer to Whisper Large-v2. $\Delta$WER = WER(attacked) $-$ WER(clean) on LibriSpeech. Attacks with higher perceptual distortion achieve higher ASR, indicating Whisper responds primarily to distortion magnitude.}
\label{tab:whisper_results}
\end{table}

\section{Layer-wise Attention Analysis}
\label{appendix:layerwise}

This section provides detailed layer-wise analysis of attention patterns under different attack conditions. Understanding how attacks affect attention at different depths of the network provides insight into the mechanisms by which adversarial perturbations induce failures.

\subsection{Summary Statistics}

Tables~\ref{tab:layerwise_audio_summary} and~\ref{tab:layerwise_video_summary} report summary statistics for attention allocation to audio and video tokens across the 28 layers of VideoLLAMA2.

\begin{table}[ht]
\centering
\adjustbox{max width=\linewidth}{
\begin{tabular}{lccc}
\toprule
\textbf{Attack Objective} & \textbf{Mean} & \textbf{Max} & \textbf{Min}\\
\midrule
$\mathcal{L}_\text{negLM}$ & 9.55 & 17.77 & 4.83 \\
$\mathcal{L}^{(\text{cos})}$ & 13.55 & 26.36 & 8.08 \\
$\mathcal{L}^{(\text{visionatt})}$ & 14.03 & 26.47 & 8.57 \\
$\mathcal{L}^{(\text{audioatt})}$ & 21.49 & 26.24 & 19.50 \\
$\mathcal{L}^{(\text{randatt})}$ & 10.51 & 20.91 & 3.36 \\
$\mathcal{L}^{(\text{hidden-cos})}$ & 13.55 & 25.06 & 5.42 \\
$\mathcal{L}^{(\text{combined})}$ & \textbf{29.23} & 30.34 & 25.96 \\
\bottomrule
\end{tabular}
}
\caption{Summary statistics for audio token attention across 28 layers of VideoLLAMA2. The combined attack produces uniformly elevated audio attention (mean 29.23) with small variance across layers.}
\label{tab:layerwise_audio_summary}
\end{table}

\begin{table}[ht]
\centering
\adjustbox{max width=\linewidth}{
\begin{tabular}{lccc}
\toprule
\textbf{Attack Objective} & \textbf{Mean} & \textbf{Max} & \textbf{Min}\\
\midrule
$\mathcal{L}_\text{negLM}$ & 14.94 & 34.64 & 6.12 \\
$\mathcal{L}^{(\text{cos})}$ & 13.96 & 30.73 & 5.82 \\
$\mathcal{L}^{(\text{visionatt})}$ & 13.49 & 29.88 & 5.30 \\
$\mathcal{L}^{(\text{audioatt})}$ & 13.85 & 29.86 & 5.49 \\
$\mathcal{L}^{(\text{randatt})}$ & 14.65 & 34.81 & 5.96 \\
$\mathcal{L}^{(\text{hidden-cos})}$ & 14.43 & 29.66 & 6.53 \\
$\mathcal{L}^{(\text{combined})}$ & 13.30 & 28.34 & 5.23 \\
\bottomrule
\end{tabular}
}
\caption{Summary statistics for video token attention across 28 layers of VideoLLAMA2. Video attention remains relatively stable across attack conditions compared to audio attention.}
\label{tab:layerwise_video_summary}
\end{table}

\subsection{Layer-wise Attention Patterns}

\begin{figure*}[ht]
    \centering
    \includegraphics[width=\linewidth]{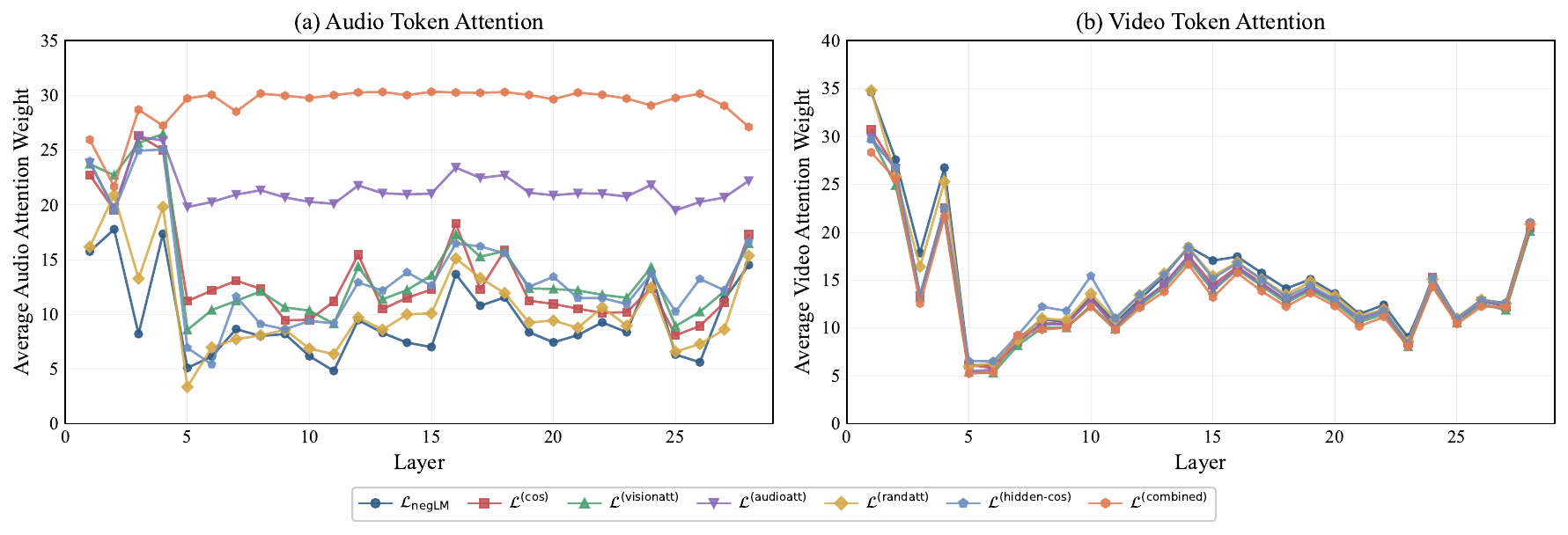}
    \caption{Layer-wise attention analysis for VideoLLAMA2. Left: Average attention mass on audio tokens per layer. Right: Average attention mass on video tokens per layer. Different attack objectives induce distinct attention patterns across the 28 transformer layers. The combined attack ($\mathcal{L}^{(\text{combined})}$) shows uniformly elevated audio attention across all layers.}
    \label{plot:layerwise_attn}
\end{figure*}

Figure~\ref{plot:layerwise_attn} visualizes the average attention mass assigned to audio and video tokens at each transformer layer under different attack objectives. Several patterns emerge from this analysis.

Video attention exhibits a characteristic front-loaded pattern across all conditions: attention is highest in early layers (1--5), then stabilizes at lower values through middle and late layers. This pattern reflects the model's processing of visual information, where early layers encode low-level features and later layers integrate them with other modalities.

Audio attention behaves differently depending on the attack objective. Under clean conditions and weak attacks ($\mathcal{L}_{\text{negLM}}$, $\mathcal{L}^{(\text{randatt})}$), audio attention remains relatively flat across layers. However, attacks that explicitly target audio pathways ($\mathcal{L}^{(\text{audioatt})}$, $\mathcal{L}^{(\text{combined})}$) induce a global upward shift in audio-token attention that is uniform across all layers. This uniform elevation suggests that these attacks modify how the model processes audio information at a fundamental level, rather than targeting specific computational stages.

The combined attack produces the most dramatic effect, with audio attention elevated to 25--30 across all layers compared to 5--17 under clean conditions. This three-fold increase in audio attention correlates with the attack's strong performance (96.03\% ASR), supporting the hypothesis that forcing the model to attend heavily to corrupted audio is an effective attack strategy.

\section{Perturbation Visualization}
\label{appendix:waveform}

\begin{figure*}[ht]
    \centering
    \includegraphics[width=\linewidth]{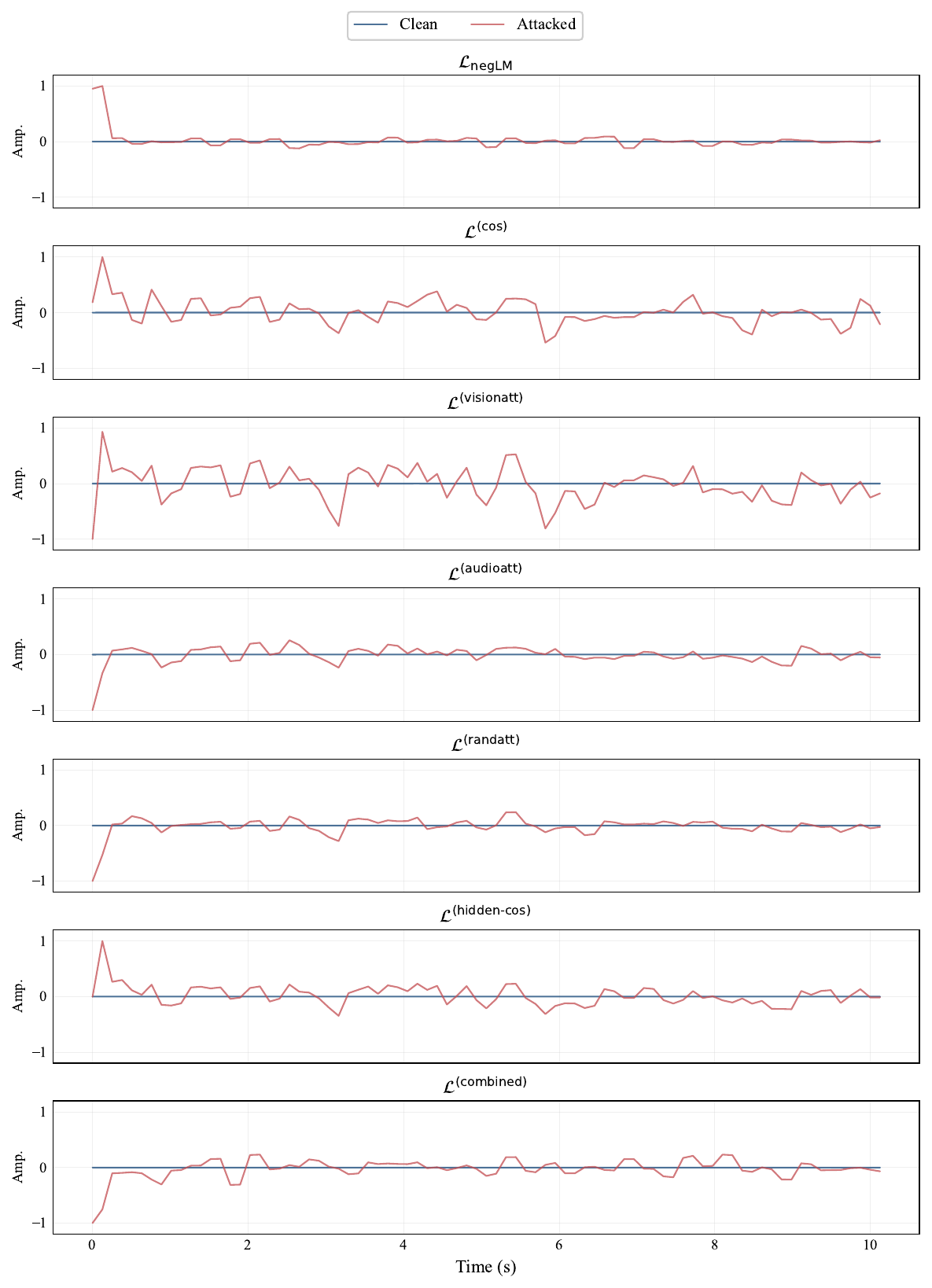}
    \caption{Waveform visualization of learned audio-only adversarial perturbations. Clean (blue) and attacked (red) audio waveforms are overlaid for perturbations optimized using different loss functions. Sampling rate is reduced for visualization clarity. Effective attacks produce perturbations that remain structurally similar to the clean signal.}
    \label{plot:clean_vs_attacked_perturbation}
\end{figure*}

Figure~\ref{plot:clean_vs_attacked_perturbation} visualizes the waveforms of adversarial audio obtained after optimizing different loss functions, overlaid on the clean reference signal. Understanding the temporal structure of learned perturbations provides insight into the attack mechanisms.

\subsection{Initialization and Structure}

All perturbations are initialized using a natural bell-ringing sound sampled from the AVQA dataset. The corresponding video is explicitly excluded from training to prevent data leakage. This initialization provides a structured starting point with temporal patterns that can be refined through optimization. The bell sound contains a mixture of transient attacks (the initial strike) and sustained resonances (the ringing decay), providing diverse spectral content.

\subsection{Perturbation Characteristics}

Across loss functions, the learned perturbations exhibit several common characteristics:

\textit{(1) Low amplitude:} Perturbations remain visually similar to the clean signal in waveform plots, with amplitudes constrained by the attack budget $\varepsilon$.

\textit{(2) Temporal smoothness:} Unlike random noise, the learned perturbations maintain temporal coherence inherited from the natural audio initialization.

\textit{(3) Structured deviation:} Effective attacks ($\mathcal{L}^{(\cos)}$, $\mathcal{L}^{(\text{combined})}$) introduce small but consistent modifications that accumulate across time, rather than isolated high-amplitude spikes.

The attention-based attacks ($\mathcal{L}^{(\text{audioatt})}$, $\mathcal{L}^{(\text{hidden-cos})}$) produce the least visible deviation from the clean signal, yet still achieve measurable attack success. This reinforces the finding that structured, low-distortion perturbations can be more effective than high-distortion approaches when targeting specific computational mechanisms.

\end{document}